\documentclass[english,twocolumn]{revtex4}
\usepackage[T1]{fontenc}
\usepackage[latin9]{inputenc}
\usepackage{verbatim}
\usepackage{float}
\usepackage{amsbsy}
\usepackage{amstext}
\usepackage{amssymb}
\usepackage{graphicx}
\usepackage{esint}

\makeatletter

\floatstyle{ruled}
\newfloat{algorithm}{tbp}{loa}
\providecommand{\algorithmname}{Algorithm}
\floatname{algorithm}{\protect\algorithmname}

\@ifundefined{textcolor}{}
{%
 \definecolor{BLACK}{gray}{0}
 \definecolor{WHITE}{gray}{1}
 \definecolor{RED}{rgb}{1,0,0}
 \definecolor{GREEN}{rgb}{0,1,0}
 \definecolor{BLUE}{rgb}{0,0,1}
 \definecolor{CYAN}{cmyk}{1,0,0,0}
 \definecolor{MAGENTA}{cmyk}{0,1,0,0}
 \definecolor{YELLOW}{cmyk}{0,0,1,0}
}

\usepackage[percent]{overpic}

\makeatother

\usepackage{babel}
\begin{document}
\global\long\def\braketop#1#2#3{\left\langle #1\vphantom{#2}\vphantom{#3}\left|#2\vphantom{#1}\vphantom{#3}\right|#3\vphantom{#1}\vphantom{#2}\right\rangle }
\global\long\def\braket#1#2{\left\langle \left.#1\vphantom{#2}\right|#2\right\rangle }
\global\long\def\bra#1{\left\langle #1\right|}
\global\long\def\ket#1{\left|#1\right\rangle }
\global\long\def\ketbra#1#2{\left|#1\vphantom{#2}\right\rangle \left\langle \vphantom{#1}#2\right|}
\global\long\def\figlab#1{(#1)}
\global\long\def\vec#1{\mathbf{#1}}
$ $

\title{Extending the Concept of Probability Flux}

\date{05/16/12}

\author{Douglas J. Mason, Mario F. Borunda, and Eric J. Heller}

\affiliation{Department of Physics, Harvard University, Cambridge, MA 02138, USA}
\begin{abstract}
We develop the Husimi map for visualizing quantum wavefunctions using
coherent states as a measurement of the local phase space to produce
a vector field related to the probability flux. Adapted from the Husimi
projection, the Husimi map is complimentary to the usual flux map,
since they are identical for small coherent states. By improving our
understanding of the flux operator and offering a robust and flexible
alternative, we show how the Husimi projection can provide a map to
the classical dynamics underlying a quantum wavefunction. We demonstrate
its capabilities on bound systems with electromagnetic fields, as
well as on open systems on and off resonance.
\end{abstract}
\maketitle

\section{Introduction}

\begin{figure}
\begin{centering}
\includegraphics[angle=-90,width=0.85\columnwidth]{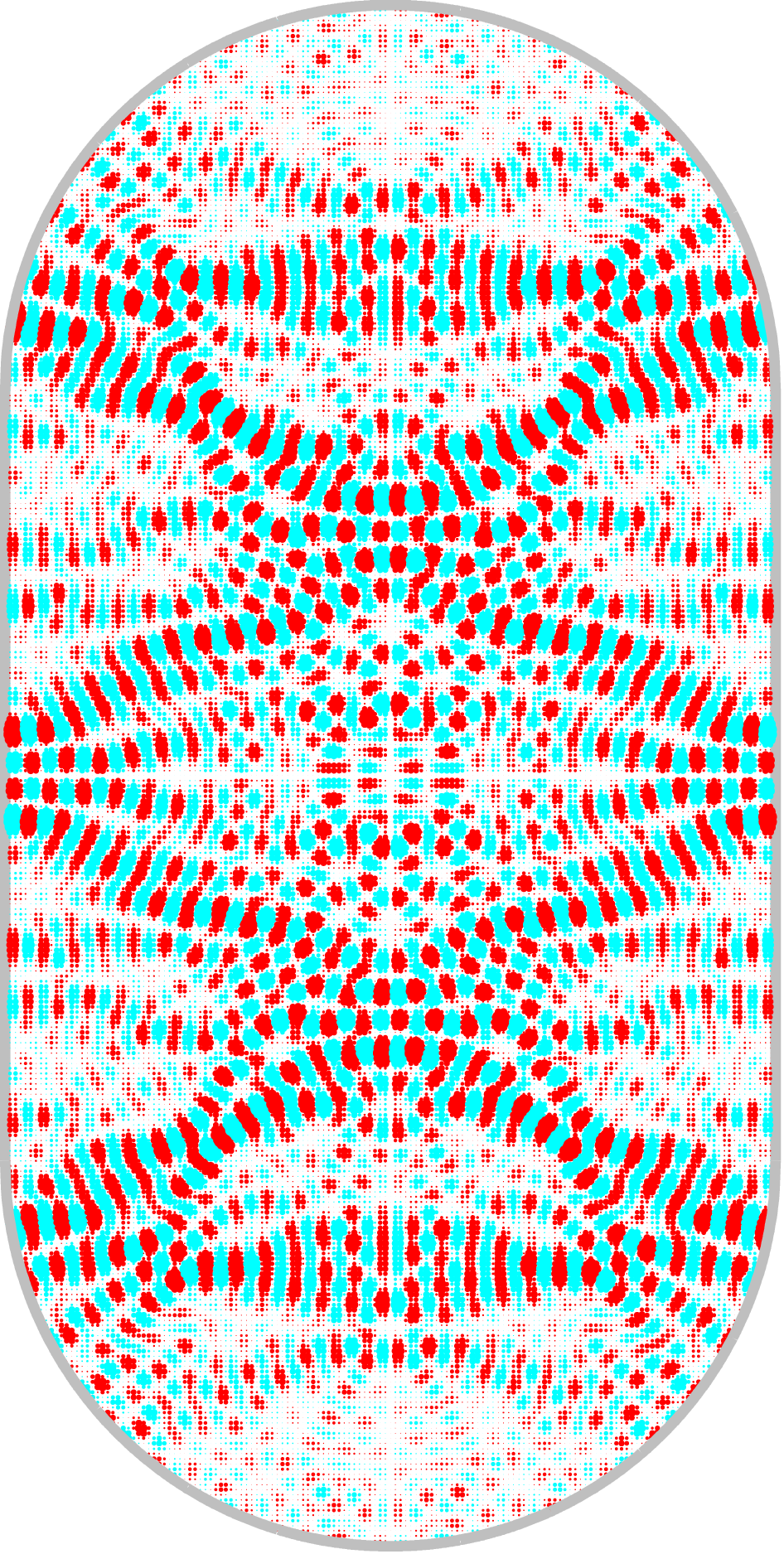}
\par\end{centering}

\caption{\label{fig:Intro}A scarred eigenstate of the stadium billiard problem
is a particle in a box with the shape shown, revealing the strong
influence of classical orbits. The traditional flux provides no help:
it is uniformly 0 inside the billiard.}
\end{figure}

The probability flux, or probability current, is introduced in quantum
mechanics textbooks as a deterministic operator that can be calculated,
but its connection to experiment is often left to the reader's imagination.
The flux operator, whose expectation over the wavefunction gives the
traditional flux $\vec j\left(\vec r,\vec p\right)$, is defined as
\begin{equation}
\hat{\mathbf{j}}_{\mathbf{r}}=\frac{1}{2m}\left(\ketbra{\mathbf{r}}{\mathbf{r}}\hat{\vec p}+\hat{\vec p}\ketbra{\vec r}{\vec r}\right),\label{eq:Flux-Operator}
\end{equation}
where $m$ is the mass of a particle in the system, and $\vec r$
and $\vec p$ the position and momentum. The concept of ``flux at
a point'' seems paradoxical because we say something about momentum
while also knowing position precisely. This raises the question: Can
the flux even be measured? 

On the other hand, probability flux vanishes on stationary states
for systems with time-reversal symmetry. This is a shame, since strong
semiclassical connections between trajectory flow and quantum eigenstates
lie completely hidden in the universal value of 0 for the flux. Consider
the example in Fig.~\ref{fig:Intro}, where the strong influence
of classical orbits is seen in the scarred eigenstate\cite{PhysRevLett.53.1515}.
For this bound system, the flux is always zero, but when it is coupled
to a continuum, flux becomes useful as a tool for examining its dynamics,
even though information about the dynamics clearly exists before the
coupling. Is it possible to bridge this gap?

By using coherent state projections, also known as Husimi projections\cite{Husimi},
we can reveal the meaning of the flux operator \emph{and }see how
to extend it to become much more useful. The experimental equivalent
of a flux map has not been discussed because it is effectively impossible
to measure -- determining the flux, even at a single point, requires
an infinite number of measurements. Instead, we offer an experimentally
feasible extension of the flux operator based on Husimi projections
which produces identical results to the traditional flux (Eq.~\ref{eq:Flux-Operator})
in the limit of infinitesimal coherent states. Because the Husimi
projection is able to work away from this limit and on a wider variety
of systems, it is able to bridge the gap between stationary and scattering
states and identify conductance pathways in large transport systems.

When many Husimi projections are sampled across a system, they produce
a Husimi \emph{map} which is a powerful tool for interpreting the
semiclassical behavior of wavefunctions. In addition to showing the
locations and directions of classical trajectories suggested by a
wavefunction, Husimi maps can also quantify how boundaries and external
fields affect those trajectories. Previous work laid out the foundation
for extending the flux operator using the Husimi projection\cite{Mason-PRL}.
In this paper, we present a complete discussion of the results summarized
there and demonstrate Husimi maps on a wider variety of systems with
and without external fields. We then show how to use Husimi maps to
interpret flux through various open devices.

\section{Measurement and the Flux Operator}

\subsection{The Gaussian Basis\label{sub:The-Gaussian-Basis}}

Several discussions connecting the flux to experimental measurement
exist in the literature\cite{measurement3,Measurement-schrod,measurement2};
here we offer an alternative view. We begin by identifying the eigenstates
of the flux operator and giving them a physical interpretation.

When discussing uncertainty, the Dirac basis implicit in Eq.~\ref{eq:Flux-Operator}
introduces singularities which we can avoid by replacing the delta
functions with the Gaussian basis defined as 

\begin{equation}
\braket{\vec r}{\vec r_{0},\sigma}=\left(\frac{1}{\sigma\sqrt{\pi/2}}\right)^{d/2}e^{-\left(\vec r-\vec r_{0}\right)^{2}/4\sigma^{2}},\label{eq:Gaussian}
\end{equation}
where $d$ is the number of dimensions in the system. The Gaussian
function becomes a delta function as $\sigma\rightarrow0$. In the
Gaussian basis, the flux operator is 
\begin{equation}
\hat{\vec j}_{\vec r_{0},\sigma}=\frac{1}{2m}\left(\ketbra{\vec r_{0},\sigma}{\vec r_{0},\sigma}\hat{\vec p}+\hat{\vec p}\ketbra{\vec r_{0},\sigma}{\vec r_{0},\sigma}\right).
\end{equation}

The eigenstates, projected onto each orthogonal spatial dimension
$i$, are obtained using the eigenvalue equation 
\begin{equation}
\hat{j}_{\vec r_{0},\sigma,i}\ket{\lambda_{\sigma,i}}=\lambda_{\sigma,i}\ket{\lambda_{\sigma,i}},\label{eq:EigenEq}
\end{equation}
which has a solution of the form 
\begin{equation}
\ket{\lambda_{\sigma,i}}=\ket{\vec r_{0},\sigma}+a\hat{p}_{i}\ket{\vec r_{0},\sigma}.
\end{equation}
Using the two equations 
\begin{equation}
\braketop{\vec r}{\hat{\vec p}}{\vec r_{0},\sigma}=i\hbar\sigma^{-2}\left(\vec r-\vec r_{0}\right)e^{-\left(\vec r-\vec r_{0}\right)^{2}/2\sigma^{2}}
\end{equation}
and
\begin{equation}
\braketop{\vec r_{0},\sigma}{\hat{\vec p}}{\vec r_{0,}\sigma}=0,
\end{equation}
we can write 
\begin{equation}
\hat{j}_{\vec r_{0},\sigma,i}\ket{\lambda_{\sigma,i}}=\frac{1}{2m}\left(a\left\langle \hat{p}_{i}^{2}\right\rangle _{\sigma}\ket{\vec r_{0},\sigma}+\hat{p}_{i}\ket{\vec r_{0},\sigma}\right).\label{eq:Explicit-EigenEq}
\end{equation}
Further, it is useful to find the conditions on $\lambda_{\sigma,i}$
that allow Eq.~\ref{eq:Explicit-EigenEq} to be written in the form
of Eq.~\ref{eq:EigenEq}, 
\begin{equation}
\lambda_{\sigma,i}=\frac{a}{2m}\left\langle \hat{p}_{i}^{2}\right\rangle _{\sigma};\lambda_{\sigma,i}=\frac{1}{2ma}.
\end{equation}

Since $\left\langle \hat{p}_{i}^{2}\right\rangle _{\sigma}=\frac{\hbar^{2}}{4\sigma^{2}}$,
the value of $a$ can be determined and from that we obtain the two
eigenvalues %
\begin{equation}
\lambda_{\sigma,i,\pm}=\pm\frac{\hbar}{4m\sigma}.\label{eq:Flux-eigenvalue}
\end{equation}
The eigenstates take the form 
\begin{equation}
\braket{\vec r}{\lambda_{\sigma,i,\pm}}=\braket{\vec r}{\vec r_{0},\sigma}\pm\frac{i}{\sigma}\vec e_{i}\cdot\left(\vec r-\vec r_{0}\right)\braket{\vec r}{\vec r_{0},\sigma},\label{eq:Flux-Eigenstate}
\end{equation}
where $\vec e_{i}$ is the unit vector along spatial dimension $i$.
Eq.~\ref{eq:Flux-Eigenstate} is a linear combination of two functions:
the Gaussian (Eq.~\ref{eq:Gaussian}) and its derivative. Projection
of a wavefunction onto the first term can be interpreted as measuring
its probability amplitude at point $\vec r_{0}$, and projection onto
the second term as measuring its derivative along the $i^{\text{th}}$
spatial dimension at the point $\vec r_{0}$. 

Because there are two eigenstates along each spatial dimension, we
show in Appendix \ref{sec:Deriving-the-Expectation} that determining
the flux expectation value on a wavefunction $\psi(\vec r)$ equates
to calculating two dot products according to
\begin{eqnarray}
\braketop{\psi}{\hat{j}_{\vec r_{0},\sigma,i}}{\psi} & = & \lambda_{\sigma,i,+}\left|\braket{\psi}{\lambda_{\sigma,i,+}}\right|^{2}\nonumber \\
 &  & +\lambda_{\sigma,i,-}\left|\braket{\psi}{\lambda_{\sigma,i,-}}\right|^{2}\label{eq:Flux-Expectation}
\end{eqnarray}
Eq.~\ref{eq:Flux-Expectation} can be rewritten as 
\begin{eqnarray}
\braketop{\psi}{\hat{j}_{\vec r_{0},\sigma,i}}{\psi} & = & \frac{i\hbar}{4m\sigma^{2}}[\braketop{\psi}{\vec e_{i}\cdot\left(\vec r-\vec r_{0}\right)}{\vec r_{0},\sigma}\braket{\psi}{\vec r_{0},\sigma}^{\ast}\nonumber \\
 &  & -\braketop{\psi}{\vec e_{i}\cdot\left(\vec r-\vec r_{0}\right)}{\vec r_{0},\sigma}^{\ast}\braket{\psi}{\vec r_{0},\sigma}].\label{eq:Flux-Expectation-2}
\end{eqnarray}

The traditional flux operator arises from the limit $\sigma\rightarrow0^{+}$,
at which point the two terms in Eq.~\ref{eq:Flux-Eigenstate} become
the delta function and its derivative, while the flux values of the
first eigenstates become 
\begin{equation}
\lim_{\sigma\rightarrow0^{+}}\lambda_{\sigma,i,\pm}=\pm\infty.
\end{equation}
In addition, there are an infinite number of other eigenstates with
zero eigenvalues (See Appendix \ref{sec:Deriving-the-Expectation}).
As a result, a single application of the flux at a particular point
in space $\hat{j}_{\vec r_{0},i}$ almost always results in zero,
but occasionally in an extremely large positive or negative value.
Traditionally, measurements of the flux correspond to the application
of the operator and averaging the results. Performing the averaging
over an infinite number of measurements results in the expression
equivalent to the textbook flux.%
{} %

\subsection{Connection to Coherent States\label{sub:Connection-to-Coherent}}

The prefactors before the Gaussian states in Eq.~\ref{eq:Flux-Eigenstate}
are related to the Taylor expansion 
\begin{equation}
e^{\pm\frac{i}{\sigma}\vec e_{i}\cdot(\vec r-\vec r_{0})}\approx1\pm\frac{i}{\sigma}\vec e_{i}\cdot(\vec r-\vec r_{0}).
\end{equation}
This suggests there may be a deep connection between the flux eigenstates
and the coherent state, defined as
\begin{equation}
\braket{\vec r}{\vec r_{0},\vec k_{0},\sigma}=\left(\frac{1}{\sigma\sqrt{\pi/2}}\right)^{d/2}e^{-\left(\vec r-\vec r_{0}\right)^{2}/4\sigma^{2}+i\vec k_{0}\cdot\vec r},
\end{equation}
which is a Gaussian envelope over a plane wave $e^{i\vec k_{0}\cdot\vec r}$.
Its inner product with a generic wavefunction $\psi\left(\vec r\right)$
is 
\begin{eqnarray}
\braket{\psi}{\vec r_{0},\vec k_{0},\sigma} & = & \left(\frac{1}{\sigma\sqrt{\pi/2}}\right)^{d/2}\nonumber \\
 &  & \times\int\psi\left(\vec r\right)e^{-\left(\vec r-\vec r_{0}\right)^{2}/4\sigma^{2}+i\vec k_{0}\cdot\vec r}d^{d}r.\label{eq:Husimi-Full}
\end{eqnarray}
Observing that the phase $e^{i\vec k_{0}\cdot\vec r_{0}}$ is arbitrary,
we can Taylor expand the exponential function in the limit of $\vec k_{0}\sigma\ll1$
to produce

\begin{eqnarray}
\braket{\vec r}{\vec r_{0},\vec k_{0},\sigma} & \approx & \left(\frac{1}{\sigma\sqrt{\pi/2}}\right)^{d/2}e^{-\left(\vec r-\vec r_{0}\right)^{2}/4\sigma^{2}}\nonumber \\
 &  & \times\left(1+i\vec k_{0}\cdot\left(\vec r-\vec r_{0}\right)\right)\\
 & \approx & \braket{\vec r}{\vec r_{0},\sigma}+i\vec k_{0}\cdot\left(\vec r-\vec r_{0}\right)\braket{\vec r}{\vec r_{0},\sigma}.\label{eq:Taylor-Expand}
\end{eqnarray}

Note that the dispersion relation for the free-particle continuum
is a circle with radius $k_{0}=\frac{\sqrt{2mE}}{\hbar}$, which is
independent of the orientation of $\vec k_{0}$. The second term in
Eq.~\ref{eq:Taylor-Expand} is thus proportional to the second term
in Eq.~\ref{eq:Flux-Eigenstate} for $\vec k_{0}$ projected along
the $i^{\text{th}}$ dimension. The similarity in form between Eq.~\ref{eq:Taylor-Expand}
and Eq.~\ref{eq:Flux-Eigenstate} allows us to relate the flux expectation
value from Eqs.~\ref{eq:Flux-Expectation} and \ref{eq:Flux-Expectation-2}
to coherent state projections as
\begin{eqnarray}
\lim_{\sigma k_{0}\rightarrow0}\braketop{\psi}{\hat{j}_{\vec r_{0},\sigma,i}}{\psi} & = & \frac{\hbar k_{0}}{4m\sigma^{2}}[\left|\braket{\psi}{\vec r_{0},k_{0}\vec e_{i},\sigma}\right|^{2}\nonumber \\
 &  & -\left|\braket{\psi}{\vec r_{0},-k_{0}\vec e_{i},\sigma}\right|^{2}].\label{eq:Flux-Hus-Corr}
\end{eqnarray}
The traditional flux \emph{vector} is constructed from the components
in each direction.

Unlike the Gaussian states appearing in the flux eigenstates in Eq.~\ref{eq:Flux-Eigenstate},
the physical meaning of coherent states is straightforward: they describe
a semiclassical particle minimizing the product of position and momentum
uncertainties. By the well-known uncertainty relation 
\begin{equation}
\Delta x\propto\frac{1}{\Delta k}\propto\sigma,
\end{equation}
taking $\sigma\rightarrow0$ results in coherent state measurements
with infinite uncertainty in $k$-space, and zero uncertainty in real
space. This is the limit where the traditional flux operates.

\subsection{Definition of the Husimi Projection\label{sub:Definition-of-the}\label{sub:Vector-Addition-in}}

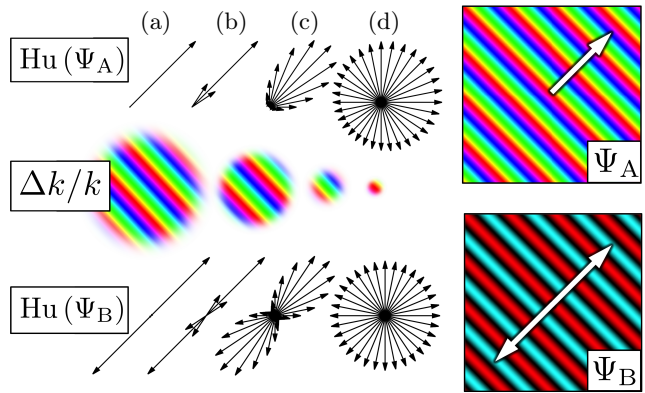
\begin{figure}
\begin{centering}
\begin{overpic}[width=1.0\columnwidth]
{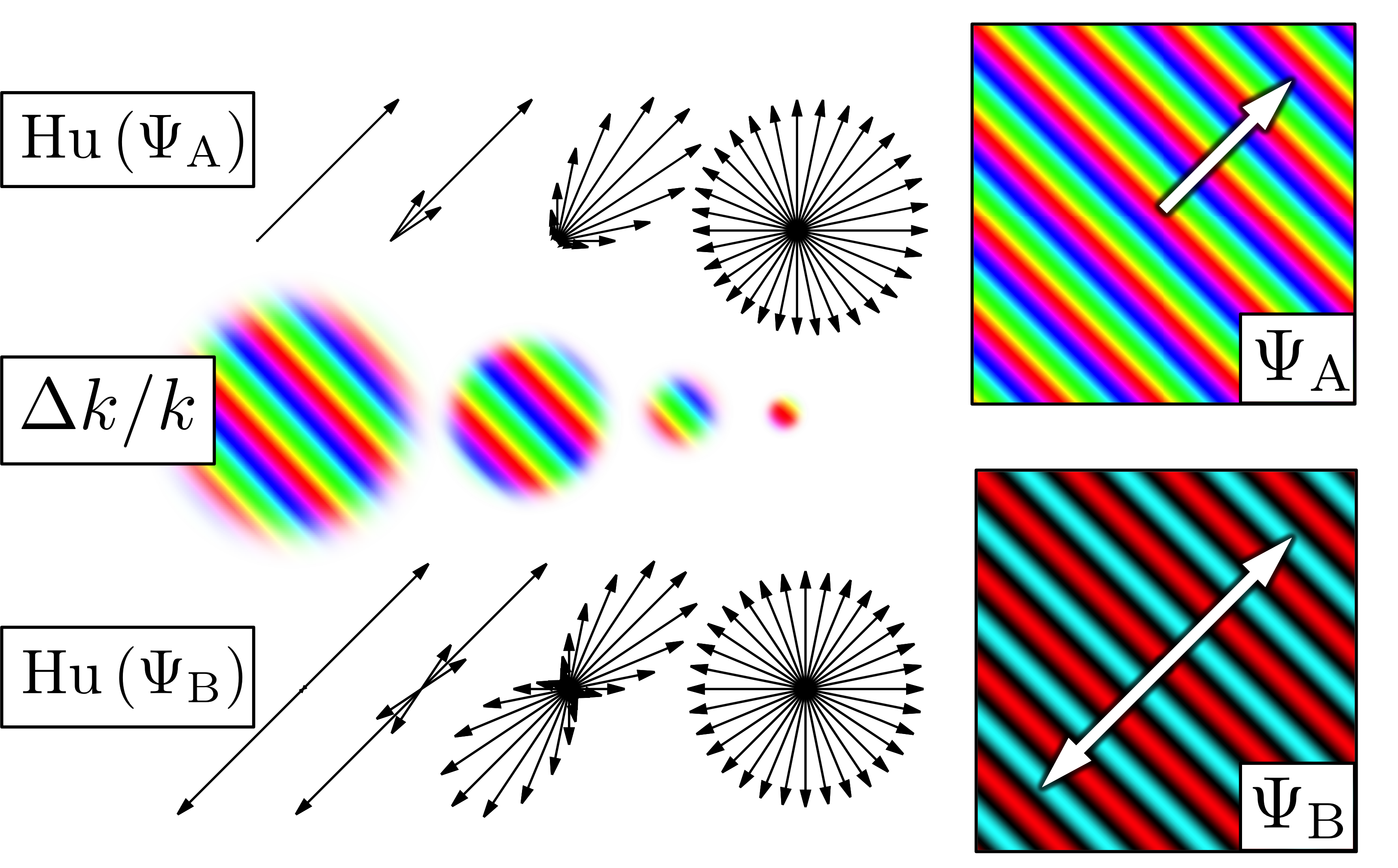}\put(20,57){\figlab{a}}\put(31.5,57)
{\figlab{b}}\put(43,57){\figlab{c}}\put(55,57){\figlab{d}}\end{overpic}
\par\end{centering}

\centering{}\caption{\label{fig:Plane Waves}Husimi vectors for 32 equally-space points
in $k$-space at left for the two wavefunctions at right: the complex
plane wave $\left(\Psi_{\text{A}}\right)$ and the cosine wave $\left(\Psi_{\text{B}}\right)$
defined in Eq.~\ref{eq:Psi}. The uncertainty for each projection
corresponds to $\Delta k/k=2\%\text{(a)},10\%\text{(b)},50\%\text{(c)},250\%\text{(d)}$,
corresponding to smaller wavepacket spreads (middle) and less distinction
between independent measurements (top and bottom). Above, we represent
the coherent wavepacket spread using schematic circles; in general,
we indicate the spread using double-arrows.}
\end{figure}

The properties of coherent states make them a suitable basis for expanding
the flux operator to a measurable definition, which we call the Husimi
function\cite{Husimi}. It is defined as a measurement of a wavefunction
$\psi(\vec r)$ by a coherent state, or ``test wavepacket'', written
as 
\begin{eqnarray}
\text{Hu}\left(\vec r_{0},\vec k_{0},\sigma;\psi(\vec r)\right) & = & \left|\braket{\psi}{\vec r_{0},\vec k_{0},\sigma}\right|^{2}.\label{eq:Husimi-Function}
\end{eqnarray}
Weighting each of these measurements by the wavevector produces a
Husimi vector; plotting all Husimi vectors at a point produces the
full Husimi projection. These projections are the sunbursts in Fig.~\ref{fig:Plane Waves},
which shows Husimi projections for the wavefunctions
\begin{eqnarray}
\Psi_{\text{A}}(\vec r) & = & e^{i\vec k_{1}\cdot\vec r}\nonumber \\
\Psi_{\text{B}}(\vec r) & = & \cos\left(\vec k_{1}\cdot\vec r\right),\label{eq:Psi}
\end{eqnarray}
where $\vec k_{1}$ points towards the upper-right. We show the wavevectors
that generate each state in the white arrow overlay.

Both wavefunctions are pure momentum states which are not spatially
localized, and constitute the building blocks for the wavefunctions
addressed in this paper. The plane wave $\Psi_{\text{A}}$ is relevant
to magnetic field states discussed in Section \ref{sub:Magnetic-Field-States}.
The cosine wave $\Psi_{\text{B}}$ corresponds to time-reversal symmetric
wavefunctions discussed in Sections \ref{sub:Circular-Well-Eigenstates}
and \ref{sub:Billiard-Eigenstates}. Both $\Psi_{\text{A}}$ \emph{and
}$\Psi_{\text{B}}$ are important for scattering wavefunctions presented
in Section \ref{sub:Sub-Threshold-Resonant-Wavefunct} which exhibit
a mixture of both properties. 

Because of the large momentum uncertainty for small $\sigma$, coherent
state projections merely reproduce the probability amplitude $\left|\psi(\vec r)\right|^{2}$
in all directions of $\vec k_{0}$, as seen in Fig.~\ref{fig:Plane Waves}d.
The flux emerges as a small residual which can be retrieved by summing
each coherent state projection weighted by $\vec k_{0}$. We call
this quantity the vector-valued Husimi flux, 
\begin{equation}
\vec{Hu}\left(\vec r_{0},\sigma;\psi(\vec r)\right)=\int\vec k_{0}\left|\braket{\psi}{\vec r_{0},\vec k_{0},\sigma}\right|^{2}d^{d}k_{0}.\label{eq:Husimi-Vector}
\end{equation}
In Appendix \ref{sub:Uncertainty-Propagation-for}, we show that as
$\sigma\rightarrow0$, the contributing points in the integral over
$k$-space reduce to just the orthogonal directions. In this limit,
we can write the Husimi flux as

\begin{eqnarray}
\lim_{\sigma\rightarrow0}\vec{Hu}\left(\vec r_{0},\sigma;\psi(\vec r)\right) & \propto & \sum_{i=1}^{d}\vec e_{i}[\left|\braket{\psi}{\vec r_{0},k_{0}\vec e_{i},\sigma}\right|^{2}\nonumber \\
 &  & -\left|\braket{\psi}{\vec r_{0},-k_{0}\vec e_{i},\sigma}\right|^{2}],\label{eq:Flux-Hus-Corr-1}
\end{eqnarray}
where $\vec e_{i}$ is the unit vector along the $i^{\text{th}}$
orthogonal direction, and we sum over $d$ dimensions. By Eq.~\ref{eq:Flux-Hus-Corr},
both sides of Eq.~\ref{eq:Flux-Hus-Corr-1} are proportional to the
traditional flux measured at point $\vec r_{0}$ so that

\begin{equation}
\braketop{\psi}{\hat{\vec j}_{\vec r_{0}}}{\psi}\propto\lim_{\sigma\rightarrow0}\vec{Hu}\left(\vec r_{0},\sigma;\psi\right).\label{eq:Husimi-Vector-Flux}
\end{equation}

For larger $\sigma$, reduced momentum uncertainty allows for substantial
variation in the coherent state projections between different directions
of $\vec k_{0}$. This can be seen in Fig.~\ref{fig:Plane Waves}
as uncertainty is reduced and uniform sunbursts (d) contract into
lobes (c), and finally to unambiguous vectors (a-b). At all uncertainties,
the absence of flux in time-reversal symmetric states can be interpreted
as the perfect cancelation of coherent state projections along each
direction in $k$-space. The equal participation of counter-propagating
flux while absent in $\Psi_{\text{A}}$ is evident in $\Psi_{\text{B}}$
as a reflected sunburst. 

The reduced momentum uncertainty for larger coherent states also reduces
spatial resolution. In the intermediate regime, we can use Husimi
projections to map the local phase space of a wavefunction. By taking
snapshots of the local phase space at many points across a system
for larger $\sigma$, we can produce a map of the classical trajectories
that correspond to a given wavefunction. These visualizations are
known as ``Husimi maps''\cite{Husimi,hellerleshouches,Husimi-map-old,Husimi-map-old2}.
Like the traditional flux map, Husimi maps can be integrated over
lines and surfaces to reveal the total probability flux current.

To produce the Husimi map, we sample Husimi projections along a grid
in spatial coordinates, since it is easier to plot, straightforward
to interpret, and allows for computing spatial derivatives (see Section
\ref{sub:Angular-Deflection}). However, other schemes may be preferred.
In Fig.~\ref{fig:Classical Paths}, for example, we sample along
classical trajectories to emphasize the quantum-classical correspondence.
While this paper addresses two-dimensional systems, Husimi projections
are equally applicable for higher-dimensional systems.

Husimi maps also have implications for experiments since they could
be measured in a fashion similar to angle-resolved photo-emission
spectroscopy (ARPES), which is currently used to measure the dispersion
relation and Fermi surfaces (for a review, see \cite{ARPES-Review}).
In the ARPES setup, a focused photon beam on a sample kicks off electrons
in the valence band. The energy of the photo-emitted electrons incorporates
both their bonding energies, which can be averaged over, and their
kinetic energy, which depends on the angle of the beam with respect
to the sample surface. 

The ARPES response function behaves similarly to coherent state projections
with $\vec k_{0}$ proportional to the beam angle. By rotating the
beam angle around the same point of intersection, the response in
different directions provides the momentum distribution of the wavefunction
at that point. Perturbations from the known dispersion relation can
then be inserted into Eq.~\ref{eq:Husimi-Vector} to obtain the flux
expectation value. 

While a narrow beam would make it possible to measure the flux vector
at the intersection point, it will be difficult to distinguish the
occasional large perturbation measurements from noise. However, wider
beams would capture additional terms from the Taylor expansion of
the coherent state in Eq.~\ref{eq:Taylor-Expand}, producing more
reliable perturbation measurements. Applying the technique at many
points across the sample would then provide the Husimi map and an
approximation to the flux map.

A question arises regarding the handling of boundaries in the system,
beyond which the wavefunction goes to zero. Our definition reduces
the magnitude of Husimi projections within distance $\sigma$ of the
boundary. When a coherent state interacts with a boundary, the boundary
can be replaced by an image wavepacket moving in the opposite direction.
In this case, reflections off the boundary amount to scattering between
wavepackets with different wavevectors. Thus, the reduction in the
Husimi projections near the boundaries is the result of wavepacket
scattering, making it possible to use Husimi maps to compute scattering
metrics along the boundary, such as angular deflection presented in
Section \ref{sub:Angular-Deflection}.

\subsection{Multi-Modal Analysis\label{sub:Multi-Modal-Analysis}}

The Husimi projections in Fig.~\ref{fig:Plane Waves} reveal that
even a single plane wave produces a range of Husimi vectors because
of the finite spread of the wavepacket. Can distinct trajectories
intersecting at a point be distinguished unambiguously? If the dominant
plane waves at a point are sufficiently separated in $k$-space, i.e.
the momentum uncertainty of the coherent state can resolve between
them, we can retrieve their wavevectors numerically using Multi-Modal
Analysis (MMA). This analytical tool can be especially useful for
time-reversal symmetric systems where both the traditional flux and
the Husimi total flux are identically zero.

\begin{figure}
\begin{centering}
\begin{overpic}[width=1.0\columnwidth]
{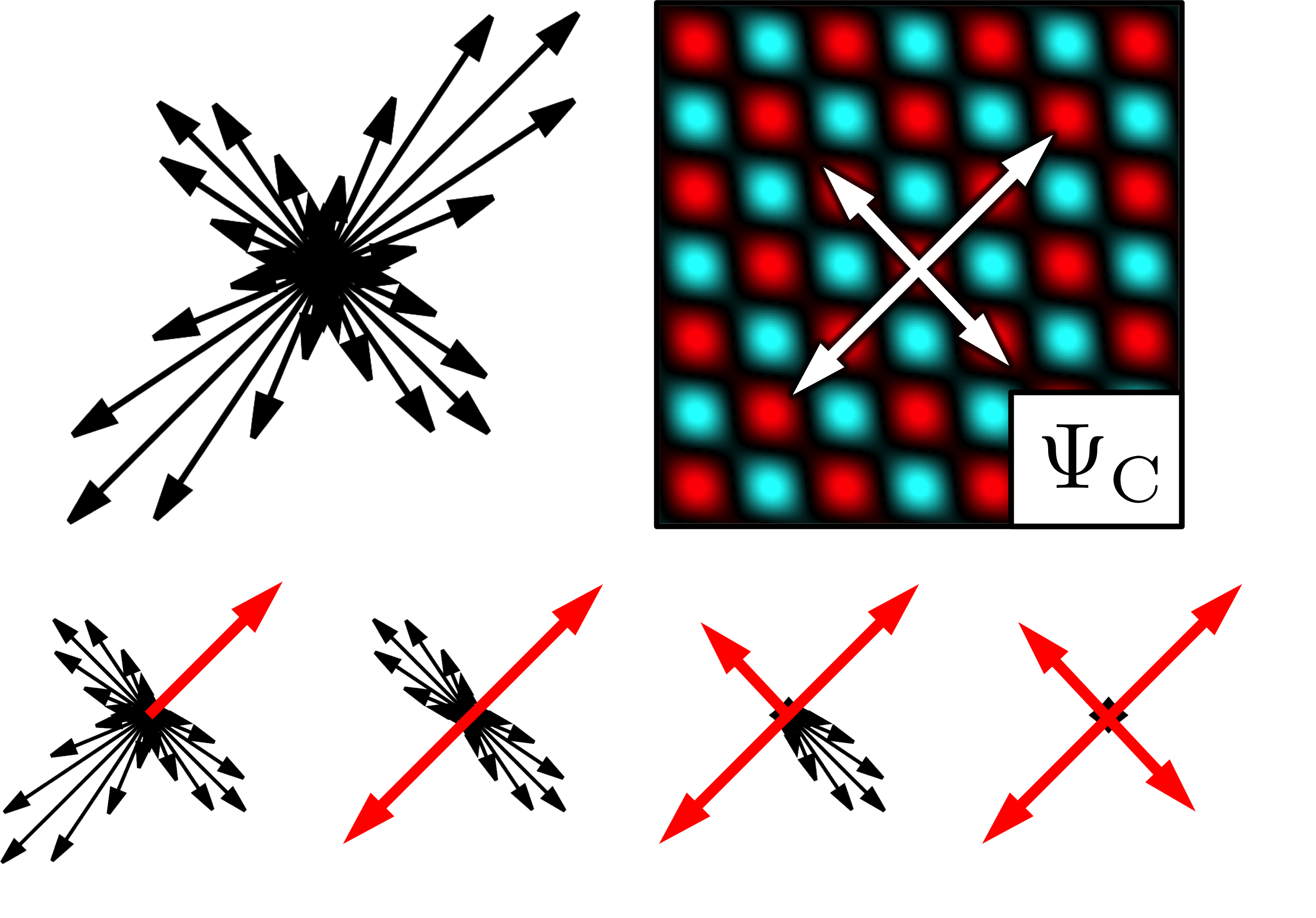}\put(6,65){\figlab{a}}\put(0,25){\figlab{b}}\put(25,25){\figlab{c}}\put(50,25){\figlab{d}}\put(74,25){\figlab{e}}\end{overpic}
\par\end{centering}

\centering{}\caption{\label{fig:Multi-Modal}Husimi vectors for 32 equally-space points
in $k$-space (a) for the double cosine waves $\left(\Psi_{\text{C}}\right)$
from Eq.~\ref{eq:Psi-2-1}. The uncertainty for each projection corresponds
to $\Delta k/k=30\%$. As the multi-modal algorithm (Algorithm \ref{alg:Multi-Modal-Analysis})
loops through each iteration (b-e), a trajectory is matched and then
subtracted from the full Husimi projection, until all major trajectories
are approximated by their appropriate values.}
\end{figure}
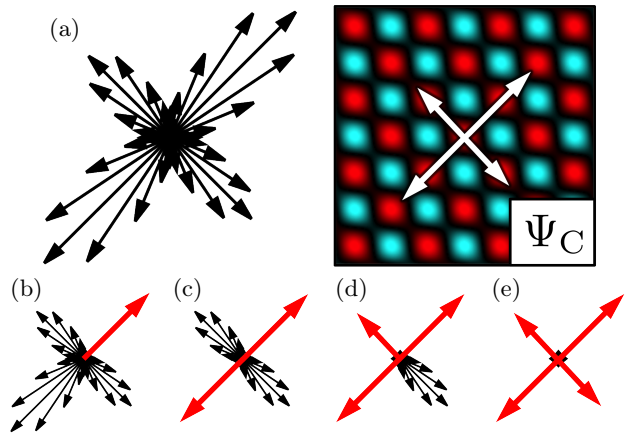

\begin{algorithm}
\begin{enumerate}
\item A set of Husimi templates on $N$ wavevectors $\left\{ \vec k_{j}\right\} $
is created for the wavefunctions $\Psi=e^{i\vec k_{i}^{\text{test}}\cdot r}$
generated by the $M$ wavevectors $\left\{ \vec k_{i}^{\text{test}}\right\} $
. Both sets of wavevectors lie along the dispersion contour. Each
template can be stored as a vector of values $\vec u_{i}$ of length
$M$ where each member corresponds to the Husimi function along the
wavevector $\vec k_{j}$.
\item Writing the Husimi projection as the vector $\vec v$, a metric is
computed $d_{i}=\vec v\cdot\vec u_{i}$ for each Husimi template.
\item The maximum of the set $\left\{ d_{i}\right\} $ is determined, and
both the wavevector $\vec k_{i}^{\text{test}}$ and the dot product
$d_{i}$ are stored.
\item The contribution of the trajectory with wavevector $\vec k_{i}^{\text{test}}$
is determined by the re-weighted vector $\vec u_{i}\frac{d_{i}}{\vec u_{i}\cdot\vec u_{i}}$.
\item The re-weighted template vector is subtracted form the projection,
that is, $\vec v\rightarrow\vec v-\vec u_{i}\frac{d_{i}}{\vec u_{i}\cdot\vec u_{i}}$.
\item All elements of $\vec v$ which are now negative are set to zero.
\item Steps 1-6 are repeated until the metric $d_{i}$ dips below a threshold.
\item The set of vectors $\left\{ d_{i}\vec k_{i}^{\text{test}}\right\} $
are used to approximate the Husimi projection
\end{enumerate}
\caption{\label{alg:Multi-Modal-Analysis}Multi-Modal Analysis (MMA)}
\end{algorithm}

Figs.~\ref{fig:Multi-Modal} demonstrates the MMA algorithm on the
pure momentum state 
\begin{eqnarray}
\Psi_{\text{C}}(\vec r) & = & \alpha\cos\left(\vec k_{1}\cdot\vec r\right)+\beta\cos\left(\vec k_{2}\cdot\vec r\right),\label{eq:Psi-2-1}
\end{eqnarray}
where $\vec k_{1}$ points towards the upper-right and $\vec k_{2}$
points towards the upper-left. We set $\alpha=1$ and $\beta=0.8$.
In Fig.~\ref{fig:Multi-Modal}a, the Husimi projection is shown with
a sizable uncertainty of $\Delta k/k=30\%$. Parts b-e iterate through
the \emph{for} loop in steps 1-6 of the MMA Algorithm. At each iteration,
the most dominant plane wave in the sunburst is modeled and then subtracted
from the projection. This is repeated until all major plane waves
have been approximated. If the dominant trajectories intersecting
at a point have sufficiently divergent momenta, not only does the
algorithm do an excellent job of modeling them, it can even compute
how many there are. In general, we stop the loop in Step 7 after a
certain number of iterations to make clearer figures.

On the other hand, when there are a number of trajectories of equal
weight whose momenta cannot be resolved by the coherent state, the
MMA Algorithm can produce unexpected results. An example of unresolved
trajectories is seen in the points sampled along the perimeter of
Fig.~\ref{fig:Ang-Mom-States}a and in the central regions of Figs.~\ref{fig:Harm-Osc}b-d
and \ref{fig:Classical Paths}. In these cases, the MMA Algorithm
approximates overlapping trajectories by first choosing their average,
and then contributing additional trajectories on either side.

When the traditional flux is non-trivial, as in Fig.~\ref{fig:Canon Trans Class}
and magnified in Fig.~\ref{fig:Flux-Multi-Modal}a, it averages over
trajectories at each point to produce the total drift flow. For this
reason the multi-modal analysis is able to augment the information
provided by the flux operator since it can show the individual trajectories
contributing to the average. For example, in Sec.~\ref{sub:Magnetic-Field-States},
we compare the drift flow highlighted by the flux to the classical
paths and the multi-modal analysis for the same system.

\subsection{Angular Deflection\label{sub:Angular-Deflection}}

The Husimi map makes it possible to compute other quantities tied
to the semiclassical underpinnings of a quantum wavefunction. This
section, and the results examined in Section \ref{sub:Billiard-Eigenstates},
focus on one: angular deflection, which reveals where system boundaries
and external fields deflect classical trajectories from straight paths
and give rise to the shape and properties of a given wavefunction.

We begin by considering the Husimi function for one point in $k$-space
measured at equally-spaced points on a grid that covers the system.
The scalar field yields a spatial map of the presence of an individual
trajectory angle, and fluctuations in the map indicate points where
classical paths deflect away from and towards the angle. Summing the
results for all wavevectors along the contour line defined by system
energy in the dispersion relation, we can derive a measurement of
angular deflection $Q_{\text{ang.}}\left(\vec r;\Psi\right)$ written
as 
\begin{equation}
Q_{\text{ang.}}\left(\vec r;\Psi\right)=\int D_{\text{abs.}}(\vec r,\vec k;\Psi)kd^{d}k.
\end{equation}
 $D_{\text{abs.}}(\vec r,\vec k;\Psi)$ is the Gaussian-weighted absolute
divergence of the Husimi map for wavevector $\vec k$ written as
\begin{eqnarray}
D_{\text{abs.}}(\vec r,\vec k;\Psi) & = & \int\sum_{i=1}^{d}\left|\frac{\text{Hu}\left(\vec k,\vec r';\Psi\right)-\text{Hu}\left(\vec k,\vec r;\Psi\right)}{\left(\vec r'-\vec r\right)\cdot\vec e_{i}}\right|\nonumber \\
 &  & \times\exp\left[\frac{\left(\vec r'-\vec r\right)^{2}}{2\sigma^{2}}\right]d^{d}r',\label{eq:Summed-Divergence}
\end{eqnarray}
where we sum over the $d$ orthogonal dimensions each associated with
unit vector $\vec e_{i}$. 

Using the Husimi map to measure angular deflection has close ties
to its initial introduction as a measurement state for building phase
diagrams\cite{Husimi}. For instance, it is possible to use the divergence
of the Husimi map for each wavevector to compute the quantum analog
of a state's Poincare map\cite{gutzwiller-chaos}. This form of the
Husimi map has been used to examine the angle of impact against a
coordinate along the boundary\cite{gutzwiller-chaos} to study chaotic
behavior in stadium billiards\cite{heller-more-husimi,Heller-billiard-with-Husimi}.

\section{Husimi Maps in Closed Systems}

\subsection{Eigenstates of the Circular System\label{sub:Circular-Well-Eigenstates}}

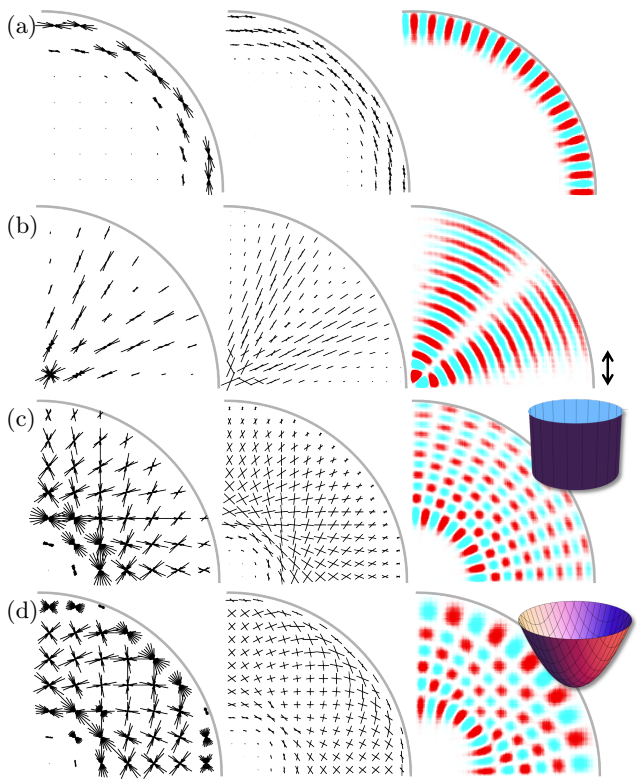
\begin{figure}
\begin{centering}

\par\end{centering}

\begin{centering}
\begin{overpic}[width=0.95\columnwidth]
{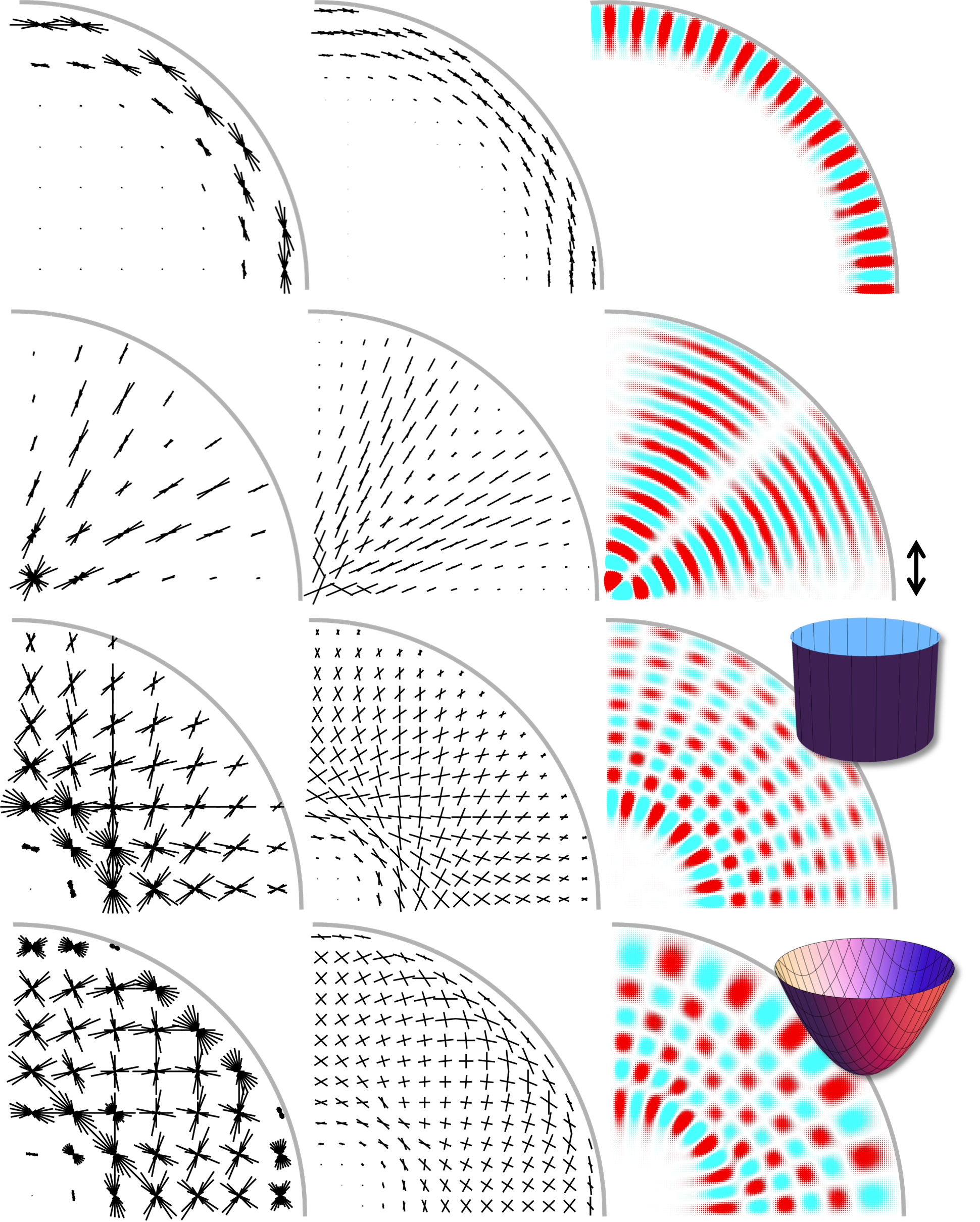}\put(-3,97){\figlab{a}}\put(-3,71){\figlab{b}}\put(-3,46){\figlab{c}}\put(-3,21){\figlab{d}}\end{overpic}
\par\end{centering}

\caption{\label{fig:Harm-Osc}\label{fig:Ang-Mom-States}Husimi maps (left),
multi-modal analysis (middle), and the wavefunction (right) for eigenstates
of the circular well (a-c) and the harmonic oscillator (d). Double-arrows
at far right indicate the spread of the coherent state which is $\Delta k/k=10\%$.
The states in (c) and (d) correspond to the classical paths in Figs.~\ref{fig:Classical Paths}a
and b respectively.}
\end{figure}
The circular well is an ideal system for demonstrating the Husimi
map since their classical dynamics are simple and can be analytically
determined. 

The Schrodinger equation can be written in radial form as 
\begin{equation}
\frac{d^{2}R(r)}{dr^{2}}+\frac{1}{r}\frac{dR(r)}{dr}+\left(k^{2}-\frac{m^{2}}{r^{2}}\right)R(r)=0.
\end{equation}
Solutions to this equation are simultaneous eigenstates of energy
and angular momentum, and thus possess the good quantum numbers $n$
(number of nodes in the radial direction) and $m$ (number of angular
nodes). Fig.~\ref{fig:Ang-Mom-States}a-c shows three such states,
the first with $n=0$, the second with $n\gg m$, and the third with
$n\approx m$. The Husimi map in each shows the clear distinction
between angular and radial components of the wavefunction, and how
they correlate with classical paths with similar properties (further
discussion of the classical correspondence can be found in Ref.~\cite{circ-well-quan-class-corr}).

To examine the harmonic oscillator state in Fig.~\ref{fig:Ang-Mom-States}d,
the Husimi projection at each point must be modified. For the circular
well, the dispersion relation is $\hbar k=\sqrt{2mE}$, but due to
the harmonic potential, it changes to $\hbar k(\vec r)=\sqrt{2m(E-V(\vec r))}$.
This means that a different sweep in $k$-space must be made at each
point to produce an accurate Husimi map. Fig.~\ref{fig:Harm-Osc}d
shows such a state with $V(\vec r)=V_{0}r^{2}$. 

The Husimi vectors in Figs.~\ref{fig:Harm-Osc}c align to suggest
straight trajectories, but the vectors in Fig.~\ref{fig:Harm-Osc}d
do not, suggesting the presence of curved paths. Moreover, projections
near the boundaries of both systems indicate that the paths of the
circular well bounce off the boundary with a consistent and acute
angle, while they graze the edge of the harmonic oscillator.

\begin{figure}
\begin{centering}

\par\end{centering}

\begin{centering}
\begin{overpic}[width=0.85\columnwidth]{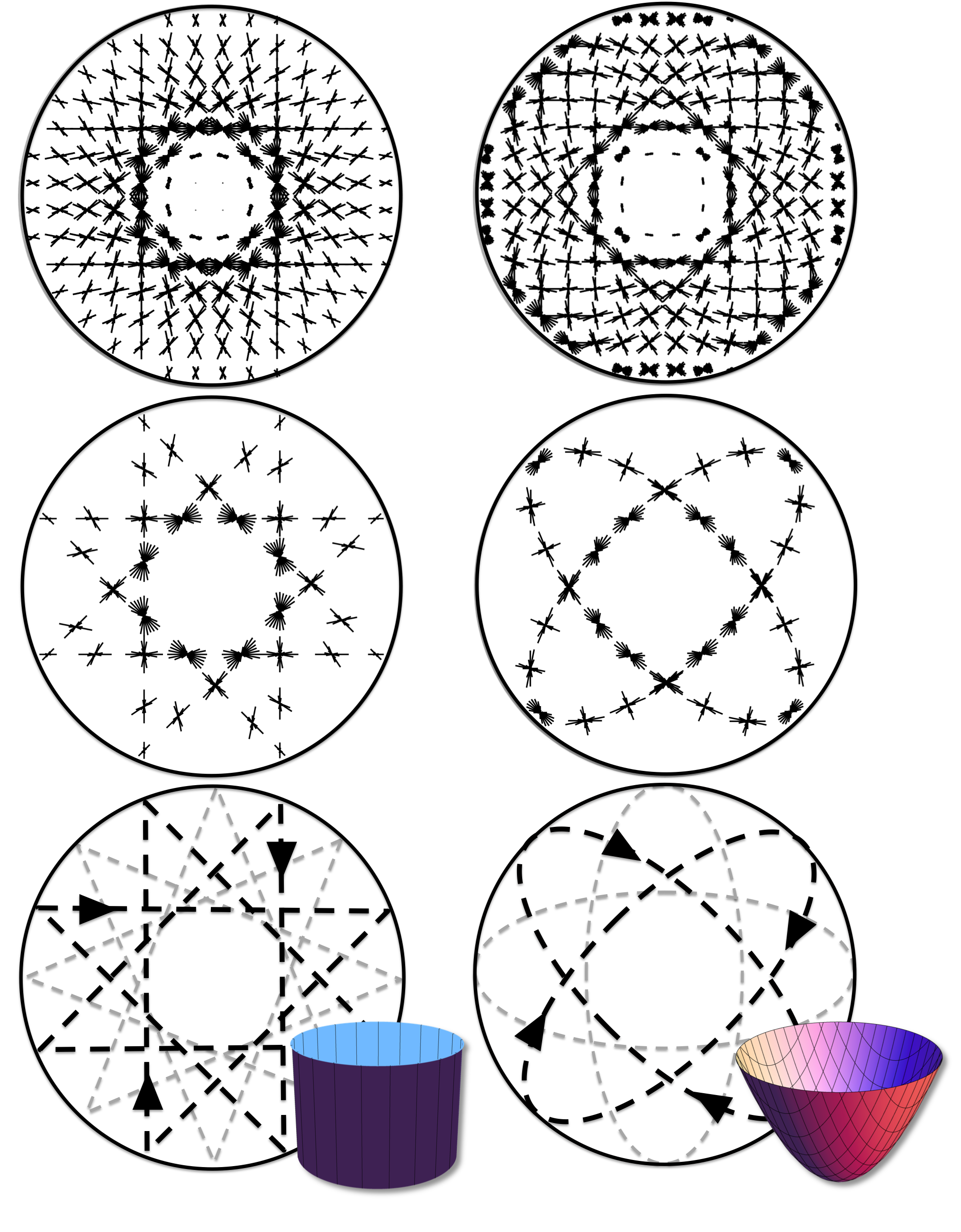}\put(2,98){\figlab{a}}\put(2,66){\figlab{b}}\put(2,34){\figlab{c}}\end{overpic}
\par\end{centering}

\caption{\label{fig:Classical Paths}Quantum-classical correspondence from
Husimi maps by sampling along classical trajectories. In part (a),
the Husimi map for the two eigenstates in Fig.~\ref{fig:Harm-Osc}c-d,
where Husimi projections are sampled along a grid. In part (b), projections
are instead sampled along classical paths that correspond to the wavefunction.
Because of rotational symmetry, however, the wavefunction is actually
created by the sum of many rotations of such paths, as indicated in
part (c).}
\end{figure}

In this paper, we have chosen to sample the Husimi projections at
equally-spaced points along a grid, which makes it possible to compute
quantities such as the angular deflection. If we instead sample along
one of the classical paths corresponding to the state, we find a set
of Husimi vectors which align themselves perfectly with the classical
path. We show these two approaches in Figs.~\ref{fig:Classical Paths}a
and \ref{fig:Classical Paths}b, which correspond to the wavefunctions
in Figs.~\ref{fig:Harm-Osc}c and \ref{fig:Harm-Osc}d respectively.

Each Husimi projection in Fig.~\ref{fig:Classical Paths}b contains
an additional set of Husimi vectors which do not align with the path.
These vectors can be understood by considering that wavefunctions
for the circular well and harmonic oscillator actually correspond
to infinitely many such paths rotated in space due to the circular
symmetry of these systems, which we indicate in Fig.~\ref{fig:Classical Paths}c.
The ``cross-hatching'' patterns in Fig.~\ref{fig:Classical Paths}a-b
arise because two rotated classical paths intersect at any point. 

Towards the center of the system, a large number of paths come into
close proximity. Even though an infinitesimal point is intersected
by only two paths, the finite spread of the coherent state is sensitive
to other paths nearby, giving rise to Husimi projections showing a
large number of trajectories with similar angles. These points in
a wavefunction can violate assumptions of the multi-modal analysis
in Section \ref{sub:Multi-Modal-Analysis} since the different trajectory
angles cannot be resolved by the finite spatial and momentum uncertainties
of each Husimi projection. As a result, the multi-modal analysis in
Figs.~\ref{fig:Harm-Osc}c and \ref{fig:Harm-Osc}d does not produce
the original paths, but their average and approximations on both sides
of the average.

\subsection{Magnetic Field\label{sub:Magnetic-Field-States}}

Systems without time-reversal symmetry can also be studied with the
Husimi technique as shown below for systems in the presence of a magnetic
field. To properly reflect these states, both the momentum operator
in Eq.~\ref{eq:Flux-Operator} and the momentum term $i\vec k_{0}\cdot\vec r_{0}$
in Eq.~\ref{eq:Husimi-Vector} must be modified to reflect the canonical
transformation
\begin{equation}
\vec p\rightarrow\vec p-q\vec A/c,
\end{equation}
where the magnetic potential $\vec A$ is defined in App.~\ref{sec:Magnetic-Fields}.%
{}

\begin{figure}
\begin{centering}
\begin{overpic}[width=0.95\columnwidth]{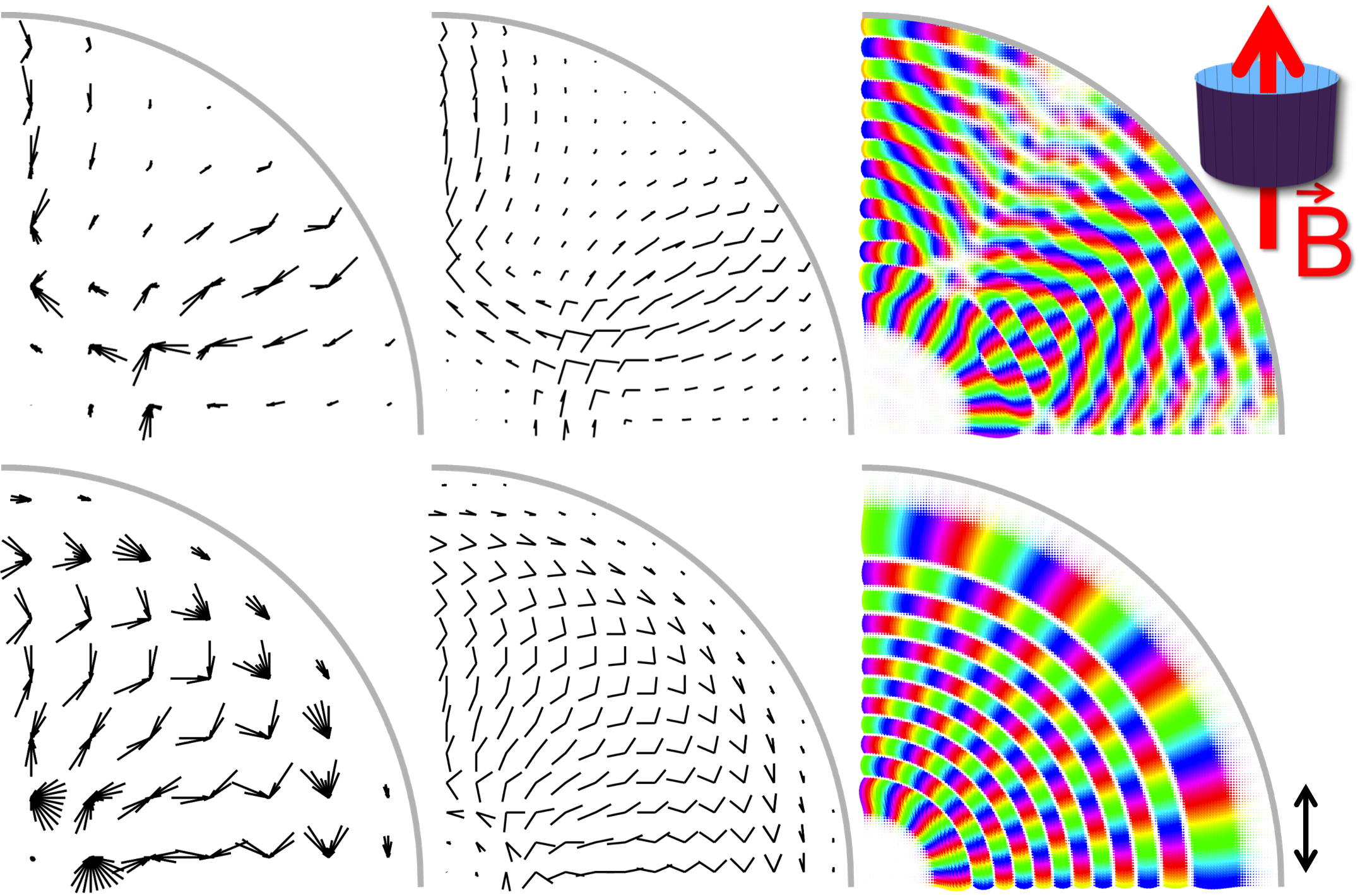}\put(-5,61){\figlab{a}}\put(-5,28){\figlab{b}}\end{overpic}
\par\end{centering}

\caption{\label{fig:Canonical Transformation}Husimi map (left), multi-modal
analysis (middle), and the wavefunction (right) for two eigenstates
of the circular well with magnetic field vectors coming out of the
plane. The magnetic field strength is set so that the cyclotron radius
is approximately $1/2$(a) and $1/3$(b) of the the system radius.
Double-arrows at far right indicate the spread of the coherent state
which is $\Delta k/k=10\%$. These states correspond to the classical
paths discussed in Fig.~\ref{fig:Canon Trans Class}. }
\end{figure}

\begin{figure}
\begin{centering}
\begin{overpic}[width=1\columnwidth]{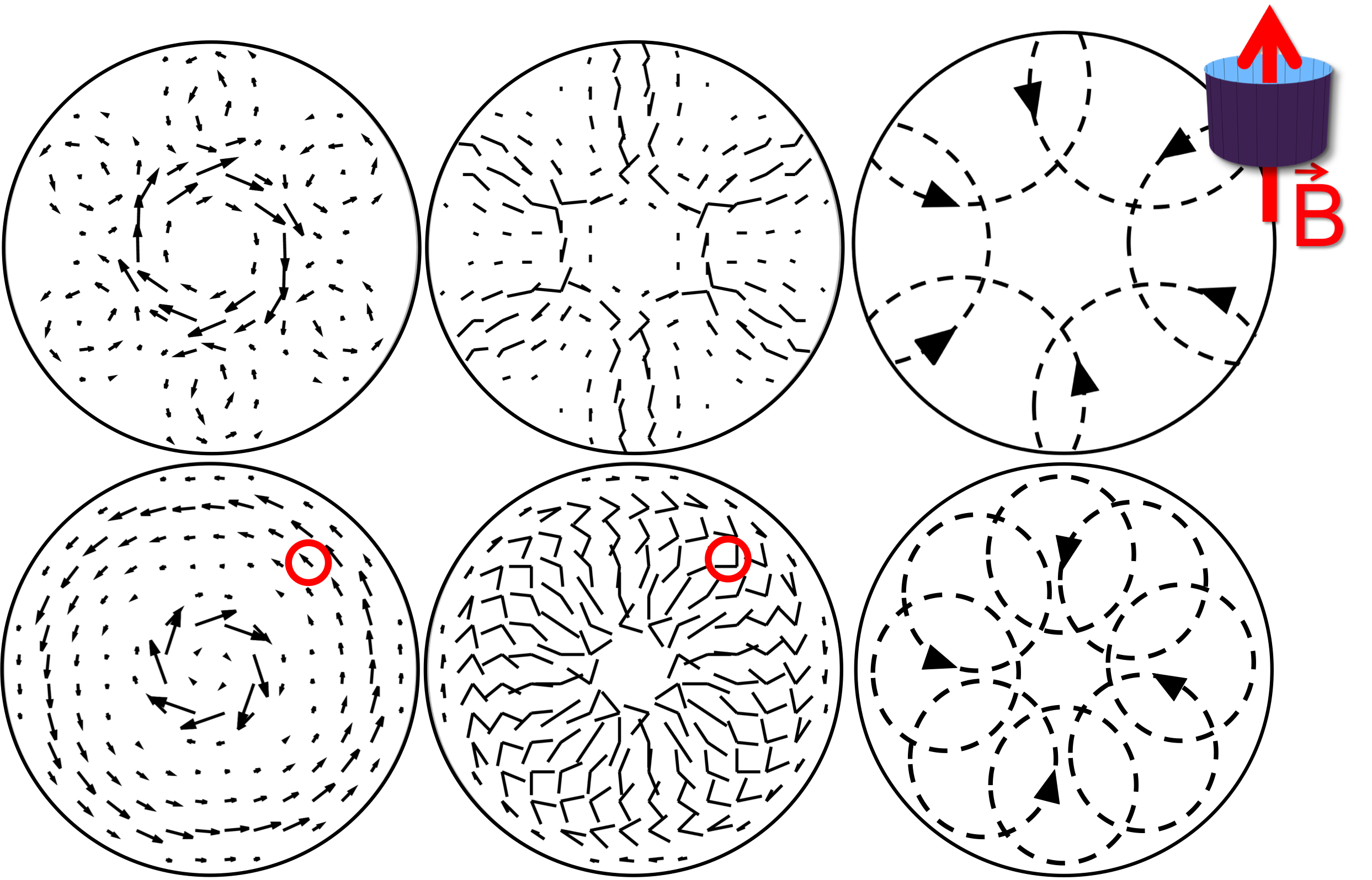}\put(0,60){\figlab{a}}\put(0,30){\figlab{b}}\end{overpic}
\par\end{centering}

\caption{\label{fig:Canon Trans Class}The flux map, multi-modal analysis,
and classical paths for the states represented in Fig.~\ref{fig:Canonical Transformation}(a-b).
The traditional flux correlates strongly with Husimi flux (Eq.~\ref{eq:Husimi-Vector})
but fails to show the classical paths suggested by the wavefunction.
Red circles correspond to magnified views in Fig.~\ref{fig:Flux-Multi-Modal}.}
\end{figure}
\begin{figure}
\begin{centering}
\begin{overpic}[width=1.0\columnwidth]
{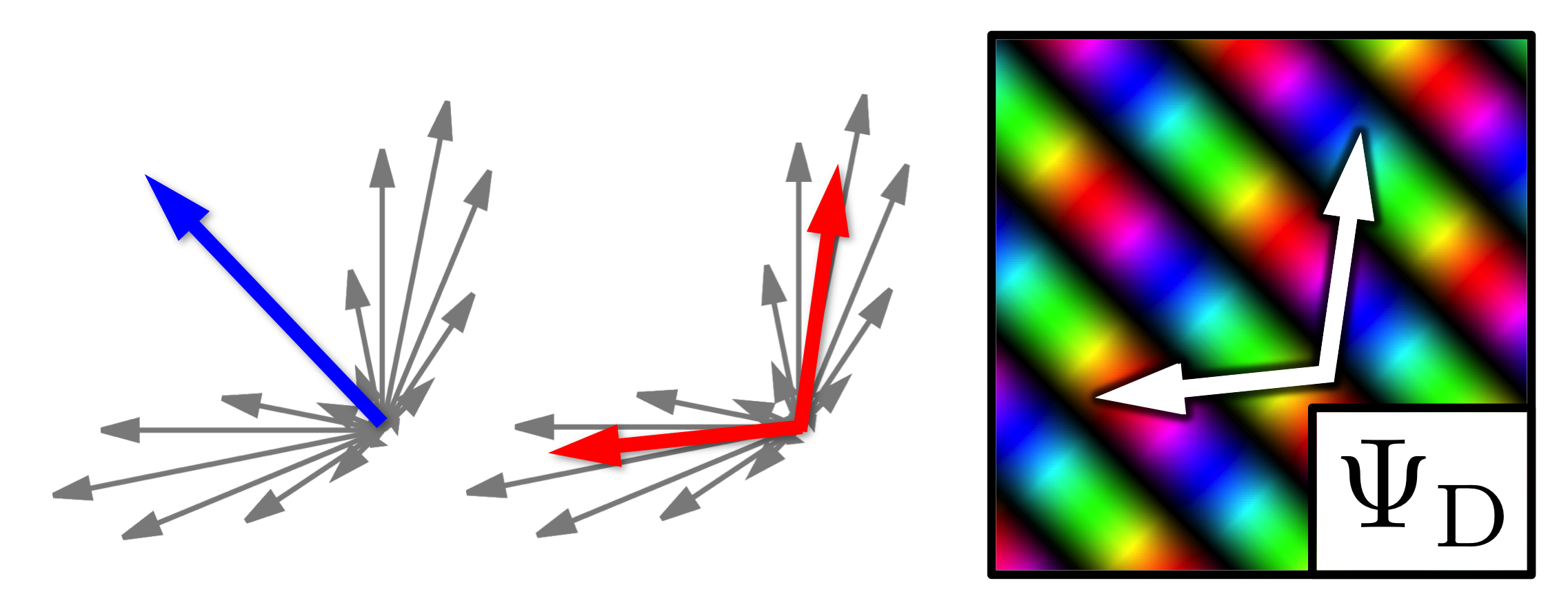}\put(3,31){\figlab{a}}\put(33,31){\figlab{b}}\end{overpic}
\par\end{centering}

\centering{}\caption{\label{fig:Flux-Multi-Modal}Husimi vectors for 32 equally-space points
in $k$-space are shown in grey for the double plane waves $\Psi_{\text{D}}$
defined in Eq.~\ref{eq:Psi-3}. The uncertainty corresponds to $\Delta k/k=30\%$.
Because of momentum uncertainty, there is spread in the Husimi vectors.
The flux operator (a,blue) averages over the Husimi vectors, while
the processed Husimi map (b,red) recovers both underlying wavevectors.
$\Psi_{\text{D}}$ is representative of the areas circled in red in
Fig.~\ref{fig:Canon Trans Class}.}
\end{figure}

Results for large magnetic fields, where the cyclotron radius is smaller
than the system size (see App.~\ref{sec:Magnetic-Fields}), are presented
in Fig.~\ref{fig:Canonical Transformation}. Unlike the circular
well states without magnetic field in Fig.~\ref{fig:Ang-Mom-States},
the wavefunctions in Fig.~\ref{fig:Canonical Transformation} do
not exhibit cross-hatching nodal patterns, but circular nodal patterns
with complex phase arguments. Projecting each phase argument onto
the real axis, however, it is easy to see that cross-hatching nodal
patterns re-emerge, suggesting the presence of multiple classical
trajectories at each point. This intuition is corroborated by the
corresponding Husimi maps for each state, which indicate circular
classical trajectories with radii corresponding to the cyclotron radius. 

In Fig.~\ref{fig:Canon Trans Class}, the full classical paths corresponding
to each state are depicted, and correlate strongly with the Husimi
map with the canonical transformation. Like the circular well states,
the presence of multiple trajectories at each point in Fig.~\ref{fig:Canon Trans Class}
can be explained by the intersection of rotated classical trajectories
that arise from rotational symmetry. For the state in Figs.~\ref{fig:Canonical Transformation}a
and \ref{fig:Canon Trans Class}a, we have artificially lifted rotational
symmetry to highlight fewer rotated paths.

Because the flux map averages over trajectories at each point, it
can often fail to indicate the full classical dynamics underlying
a quantum wavefunction. The left column of Fig.~\ref{fig:Canon Trans Class},
which shows the flux map, integrated with a Gaussian kernel corresponding
to the coherent state used to generate the Husimi map, is consequently
unable to represent the classical paths (right column), but instead
measures and total drift flow which might be the desired quantity
in some circumstances. In contrast, the multi-modal analysis in the
middle column indicates the classical paths with remarkable fidelity. 

We can appreciate the difference in detail by examining the areas
circled in red in Fig.~\ref{fig:Canon Trans Class}. In Fig.~\ref{fig:Flux-Multi-Modal},
magnified views from the flux operator, multi-modal analysis, and
full Husimi map corresponding to these ares are shown. We model this
point in the wavefunction according to the pure momentum state 
\begin{equation}
\Psi_{\text{D}}\left(\vec r\right)=e^{i\vec k_{3}\cdot\vec r}+e^{i\vec k_{4}\cdot\vec r},\label{eq:Psi-3}
\end{equation}
where $\vec k_{3}$ and $\vec k_{4}$ are indicated by the white arrows.
Not only does the flux average the full Husimi projection, it also
averages over the trajectories inferred by multi-modal analysis. As
a result, the flux map in the left column Fig.~\ref{fig:Canon Trans Class}
can be deduced from the multi-modal analysis in the middle column
by simply summing the vectors at each point.

\subsection{Stadium Billiard Eigenstates\label{sub:Billiard-Eigenstates}}

\begin{figure}
\begin{centering}
\begin{overpic}[width=0.95\columnwidth]{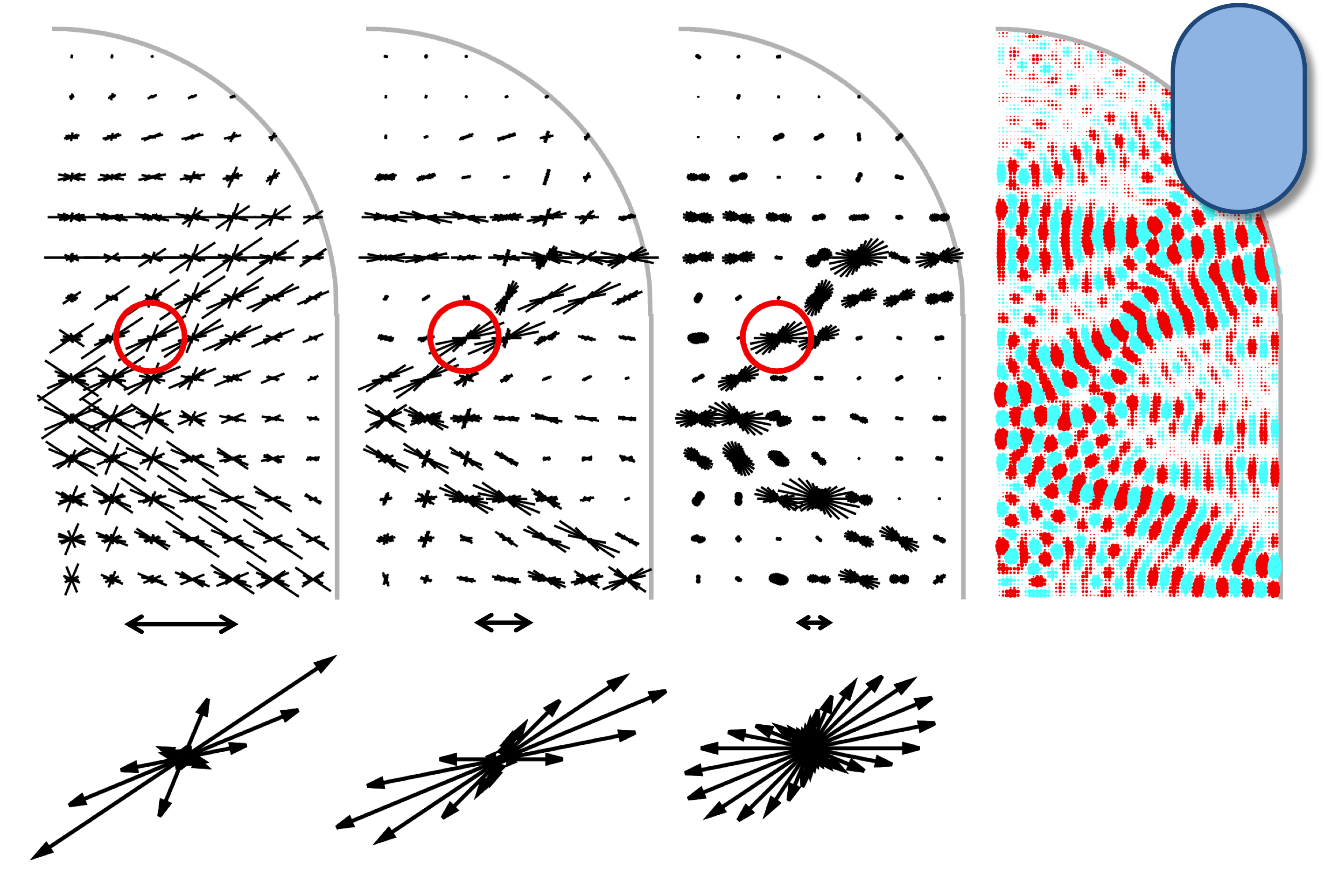}\put(-1,63){\figlab{a}}\put(22,63){\figlab{b}}\put(46,63){\figlab{c}}\put(70,63){\figlab{d}}\end{overpic}
\par\end{centering}

\caption{\label{fig:Diff Sized Gaussians}Husimi maps for the scarred stadium
billiard eigenstate. Each map uses a different spread of the measurement
wavepacket. The spread is indicated by the double-arrows on the bottom,
with relative uncertainties of $\Delta k/k=5\%$(a), $20\%$(b), and
$50\%$(c). A single Husimi projection, circled in red, is magnified
at the bottom of each representation. }
\end{figure}
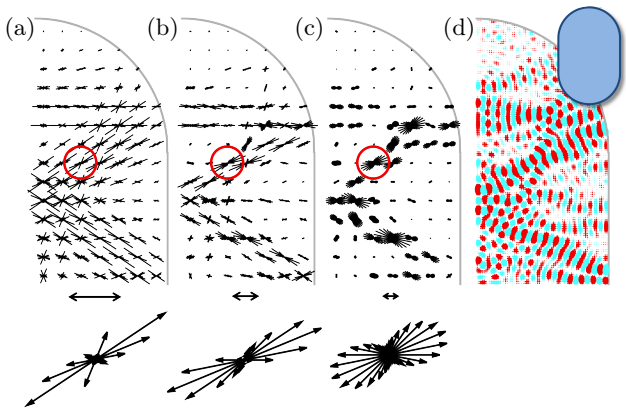

The classical dynamics of the circular stadium are integrable while
those of the Bunimovich stadium\cite{Bunimovich} are chaotic. As
a result, the stadium has been featured in many studies of \textquotedbl{}quantum
chaology\textquotedbl{}\cite{chaology0,Heller-billiard4,chaology2,Heller-billiard,Heller-billiard2,billard-scars,Heller-billiard5,gutzwiller-chaos}. 

Fig.~\ref{fig:Diff Sized Gaussians} shows three Husimi maps for
a billiard eigenstate. The wavelength at the energy of the eigenstate
is much shorter than the size of the system, allowing well-defined
scars to form, which are spawned by modestly unstable and infinitely
rare (among all the chaotic orbits) classical periodic orbits\cite{PhysRevLett.53.1515}.

For Fig.~\ref{fig:Diff Sized Gaussians}a, an extended coherent state
is used to generate the Husimi map, so that many fine features of
the wavefunction are washed out. Only the scar path (seen as a rotated
``v'' pattern in the depiction) is clearly visible. The sharply
peaked Husimi sunburst reflects both the low momentum uncertainty
of the Gaussian used and the strong dominance of the periodic orbit
pathway in the eigenfunction.

Compare this to the Husimi map in Fig.~\ref{fig:Diff Sized Gaussians}c
which is generated by a small coherent state with larger momentum
uncertainty. Here, each Husimi projection is more ambiguous, and local
variations in the wavefunction probability amplitude have a large
impact on the representation since they are no longer smoothed over.
As a result, the trajectories implied by the map no longer continue
from one projection to its neighbors and appear somewhat irregular.
In general, a compromise can be made by choosing an intermediate momentum
uncertainty, as shown in the Husimi map presented in Fig.~\ref{fig:Diff Sized Gaussians}b.
Trajectories are fairly well-resolved, and local variations are easy
to follow. Coherent states of this size provide the clearest representation
of semiclassical paths.

\begin{figure}
\begin{centering}
\begin{overpic}[width=0.95\columnwidth]{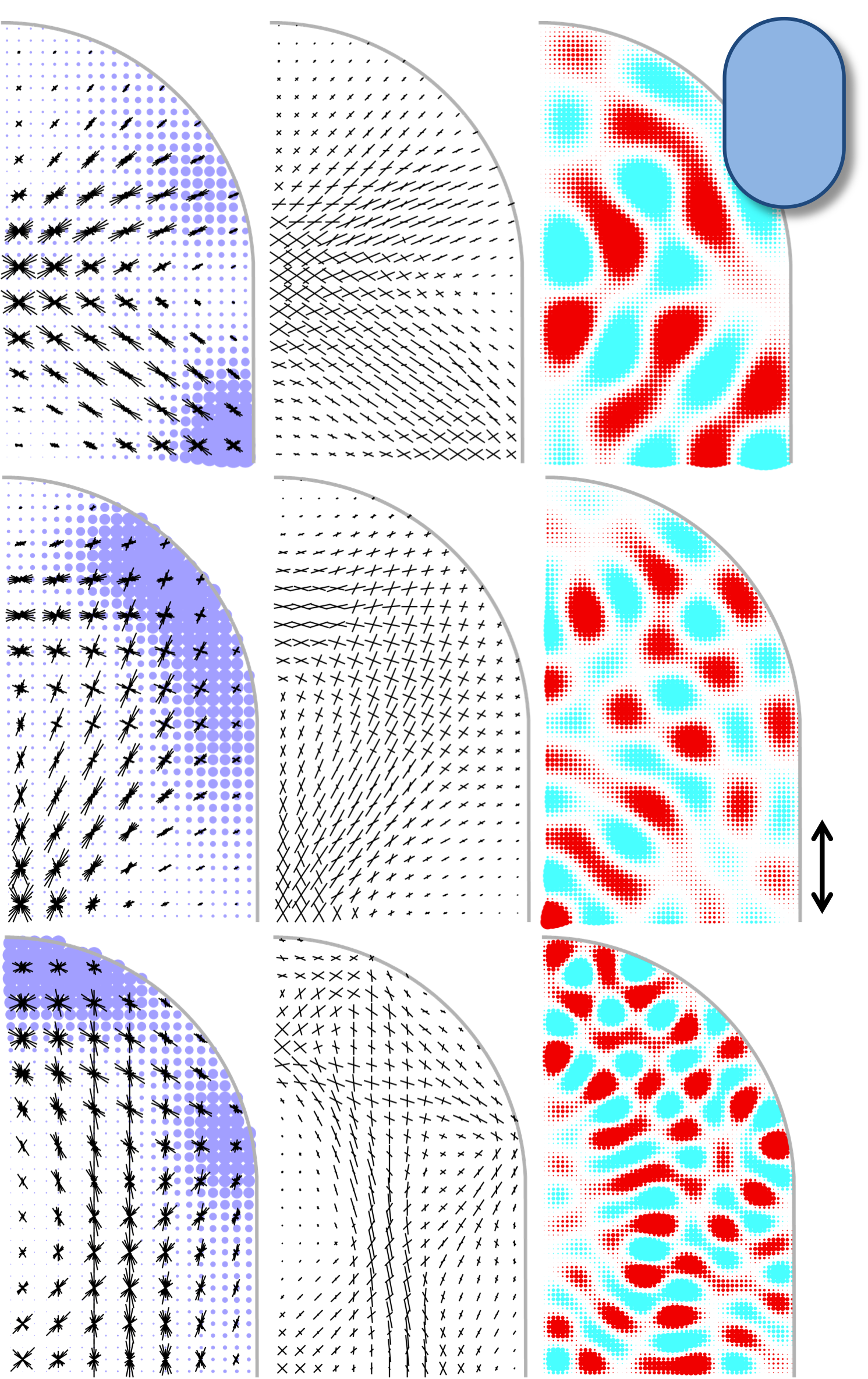}\put(-3,96){\figlab{a}}\put(-3,63){\figlab{b}}\put(-3,30){\figlab{c}}\end{overpic}
\par\end{centering}

\caption{\label{fig:Billiard Eigenstates}Three eigenstates of the stadium
billiard system with Dirichlet boundary conditions at three increasing
energies, with the Husimi map (left), multi-modal analysis (middle)
and wavefunction (right). Angular deflection is indicated in blue,
and the double arrows indicate the test wavepacket spread of $\Delta k/k=20\%$(a),
$15\%$(b) and $10\%$(c). }
\end{figure}

Even at low energies, where the wavelength is comparable to the size
of the system, stadium billiards provide another perspective on the
utility of the Husimi map. Unlike the circular system, in which the
trajectories adding up at a particular point are fairly regular and
predictable, any point in a stadium billiard eigenstate is rife with
many unpredictable trajectories. Thus, the Husimi map is an ideal
tool for lifting the veil on the underlying classical dynamics.

In Fig.~\ref{fig:Billiard Eigenstates}, we show Husimi maps for
three eigenstates of the closed stadium billiard Hamiltonian. For
each calculation, the size of the coherent state is kept constant,
but because the energy of the eigenstates increases from top-to-bottom,
the momentum uncertainty for each Husimi projection also increases.
This is reflected in the clarity of the suggested classical paths
at higher energy as well as the reduction of angular deflection in
the bulk (which acquires small positive values in the top figure due
to uncertainty, not because there is actual deflection at these points).

To the unaided eye, the wavefunctions in Fig.~\ref{fig:Billiard Eigenstates}
do not appear to emphasize isolated classical trajectories like the
high-energy stadium state in Fig.~\ref{fig:Diff Sized Gaussians},
especially since at such low energies the system only accommodates
a few wavelengths along its diameter. In the Husimi map, however,
it is quite clear that a very limited set of classical trajectories
are largely responsible for these wavefunctions, suggesting that Husimi
projections could be used to study the properties of low-energy scar
states\cite{PhysRevLett.53.1515}.

Points with high angular deflection show which parts of the system
boundary are responsible for the creation of each state, and indicate
where adiabatic changes in the boundary conditions are most likely
to affect the state\cite{Deformation-Chaotic-Billiard,Deformation-Cavity}.
This can be imagined as a quantum force on the boundary. Because
the size of the coherent state used to generate each Husimi map is
kept constant, the angular deflection penetrates into the bulk to
the same extent for each state. However, the locations of high angular
deflection along the boundary form a unique fingerprint for each state.

\section{Flux Through Open Systems}

\subsection{Sub-Threshold Resonance\label{sub:Sub-Threshold-Resonant-Wavefunct}}

\begin{figure}
\begin{centering}
\includegraphics[width=0.85\columnwidth]{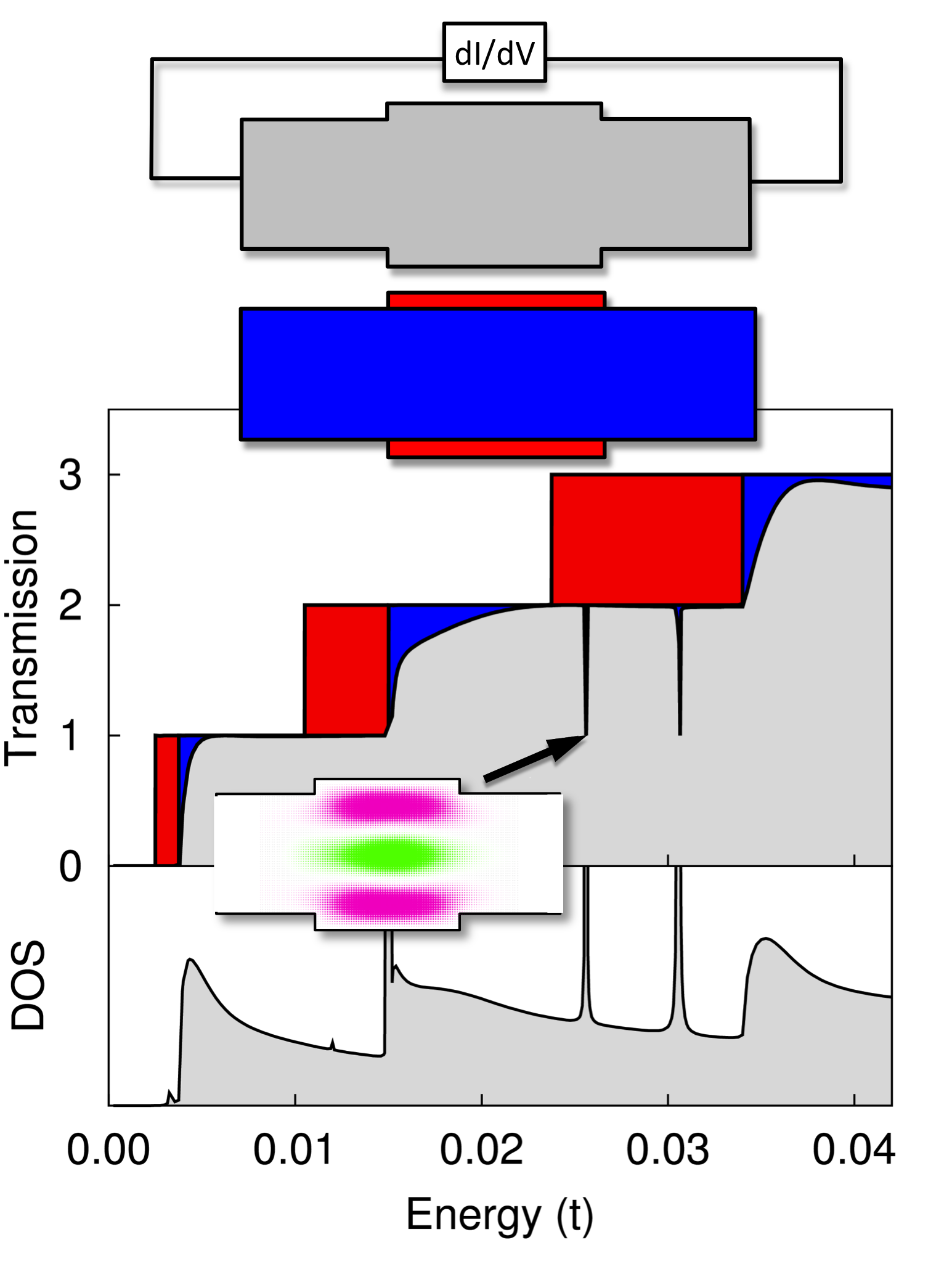}
\par\end{centering}

\caption{\label{fig:Sub-Threshold Resonance}Top: An infinite waveguide schematic
with a slight bulge in the middle (grey). This can be modeled as two
waveguides of different widths (blue and red). Bottom: In an infinite
waveguide, the transmission curve has a series of plateaus as each
transverse mode opens up (blue transmission curve). In a wider waveguide,
each mode opens up at lower energies (red curve). If only a small
segment of the waveguide is widened, then sub-threhold resonances
occur in between the energies of the narrow and wide waveguides (grey
transmission curve). These correlate with sub-threshold resonant states
which peak in the density of states (DOS) at those energies (grey
curve). Energy is given in units of $t$ where $4t$ is the band edge.}
\end{figure}

The previous section used the Husimi map to examine the semiclassical
dynamics of closed systems directly from their wavefunctions, providing
substantial benefits over the usual flux operator, which vanishes
for time reversal symmetric systems, and averages all trajectories
(thus missing criss-crossing trajectory paths, see Fig.~\ref{fig:Canon Trans Class})
for a magnetic field present. Moreover, the spread of the coherent
state used to generate the Husimi map gives it the flexibility to
examine dynamics at a variety of scales, while the flux operator is
confined to the limit of infinitesimal spread. In its traditional
guise (Eq.~\ref{eq:Flux-Operator}), the flux operator is most often
employed in scattering problems which arise when a closed system is
coupled to an environment. Is it possible to connect the semiclassical
dynamics of the closed system to the open system using the extended
Husimi flux?

In this section, we demonstrate how the Husimi flux can help interpret
the traditional flux and deepen our understanding of transport across
a device. We consider sub-threshold resonance for a waveguide that
is slightly widened along a short section (see inset, Fig.~\ref{fig:Sub-Threshold Resonance}).%

In an unperturbed waveguide, transport occurs through transverse modes
which open for transport when the system energy exceeds the transverse
energy of the mode. At these energies, the transmission function exhibits
distinct plateaus as seen in Fig.~\ref{fig:Sub-Threshold Resonance},
where the plot of the transmission for a wide(narrow) waveguide is
presented in red(blue). 

If a small section of a narrow waveguide is widened, the transverse
energy of each mode diminishes in the wider section. Thus, for each
mode, there is a range of energies bounded above by its transverse
energy in the unperturbed waveguide, and below by its energy in the
wider region. In this energy range, the mode can reside in the wider
region but cannot propagate through the narrower leads where it is
an evanescent wave. This forces it into a quasi-bound state which
is trapped in the wider region and is only weakly coupled to the environment,
causing a striking peak in the density of states, commonly known as
a Feshbach resonance\cite{Feshbach1958357}. In the quasi-bound state,
the particle bounces vertically between the walls of the perturbed
region and is unlikely to escape.

At certain energies, a particle propagating in a lower energy mode
corresponding to the narrow section interacts with the wider region
and becomes trapped in the quasi-bound state. This causes the quasi-bound
state to hybridize with the propagating mode and interfere with the
transmission in the device, as seen in Fig.~\ref{fig:Sub-Threshold Resonance}.
The suppression of transmission appears as a pair of sharp dips, accounting
for symmetric and antisymmetric versions of the Feshbach resonance.
Since the resulting wavefunction is the hybridized state which inhabits
the system at resonance, we refer to it as the resonant state.

\begin{figure}
\begin{centering}
\begin{overpic}[width=0.95\columnwidth]{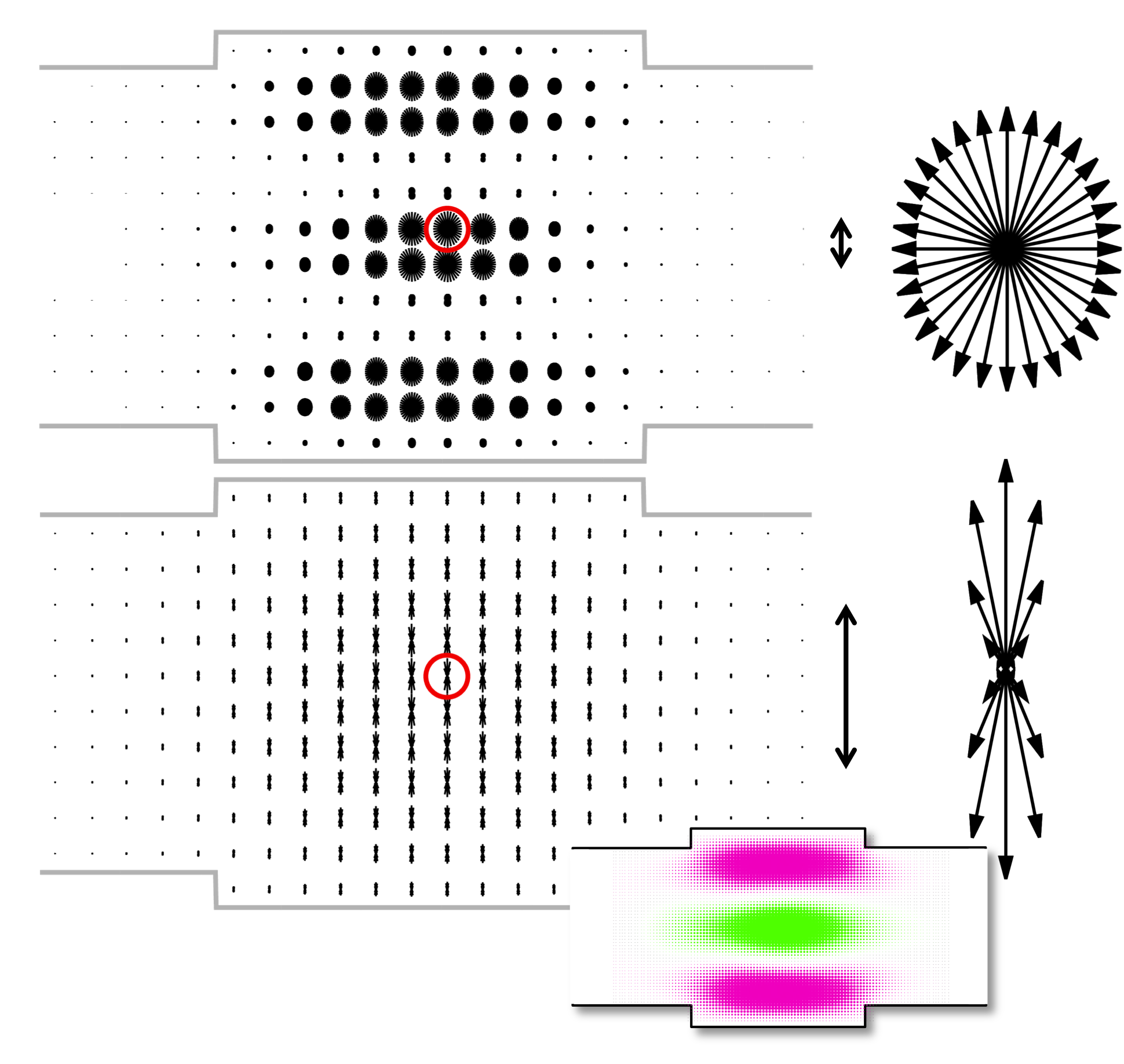}\put(2,91.5){\figlab{a}}\put(2,51.5){\figlab{b}}\end{overpic}
\par\end{centering}

\caption{\label{fig:Full-Husimis}The full Husimi map for the resonant state
(see inset) is plotted with $\Delta k/k=100\%$ (a) and $\Delta k/k=20\%$
(b). The spread of the test wavepacket is indicated by double-arrows.
A single Husimi projection (circled in red) for each map is magnified
at right. The vector sums of each map are shown in Figs.~\ref{fig:Above-At-Below}b.}
\end{figure}

We compute the wavefunction of the resonant state corresponding to
the first transmission dip in Fig.~\ref{fig:Sub-Threshold Resonance}
(indicated by the arrow in the transmission function) according to
Appendix \ref{sec:Calculating-the-Wavefunctions}. Our method allows
us to extract the pure resonant state without the second-lowest propagating
mode, which is also present at these energies. Fig.~\ref{fig:Full-Husimis}
shows the full Husimi map for this wavefunction, using coherent states
with uncertainties of $\Delta k/k=100\%$ (a) and $20\%$ (b). The
individual projections correspond strongly to the cosine-wave projections
in Fig.~\ref{fig:Plane Waves}. Spatial variations in the Husimi
map decrease as the size of the coherent state increases, as in Fig.~\ref{fig:Diff Sized Gaussians}.

The full Husimi map is indistinguishable from the quasi-bound state
and the resonant state, which is expected since the resonant state
only slightly perturbed by the propagating mode. The flux of the quasi-bound
state is zero, but exhibits characteristic vortices in the resonant
state. Moreover, as the energy is increased across resonance, the
wavefunction doesn't substantially change in appearance, while the
flux patterns alter dramatically. At first these behaviors appear
to contradict the Husimi map, but we can show that the flux patterns
correlate with subtle changes in the Husimi maps which we can retrieve
by adding all their vectors. 

\begin{figure}
\begin{centering}
\begin{overpic}[width=0.85\columnwidth]{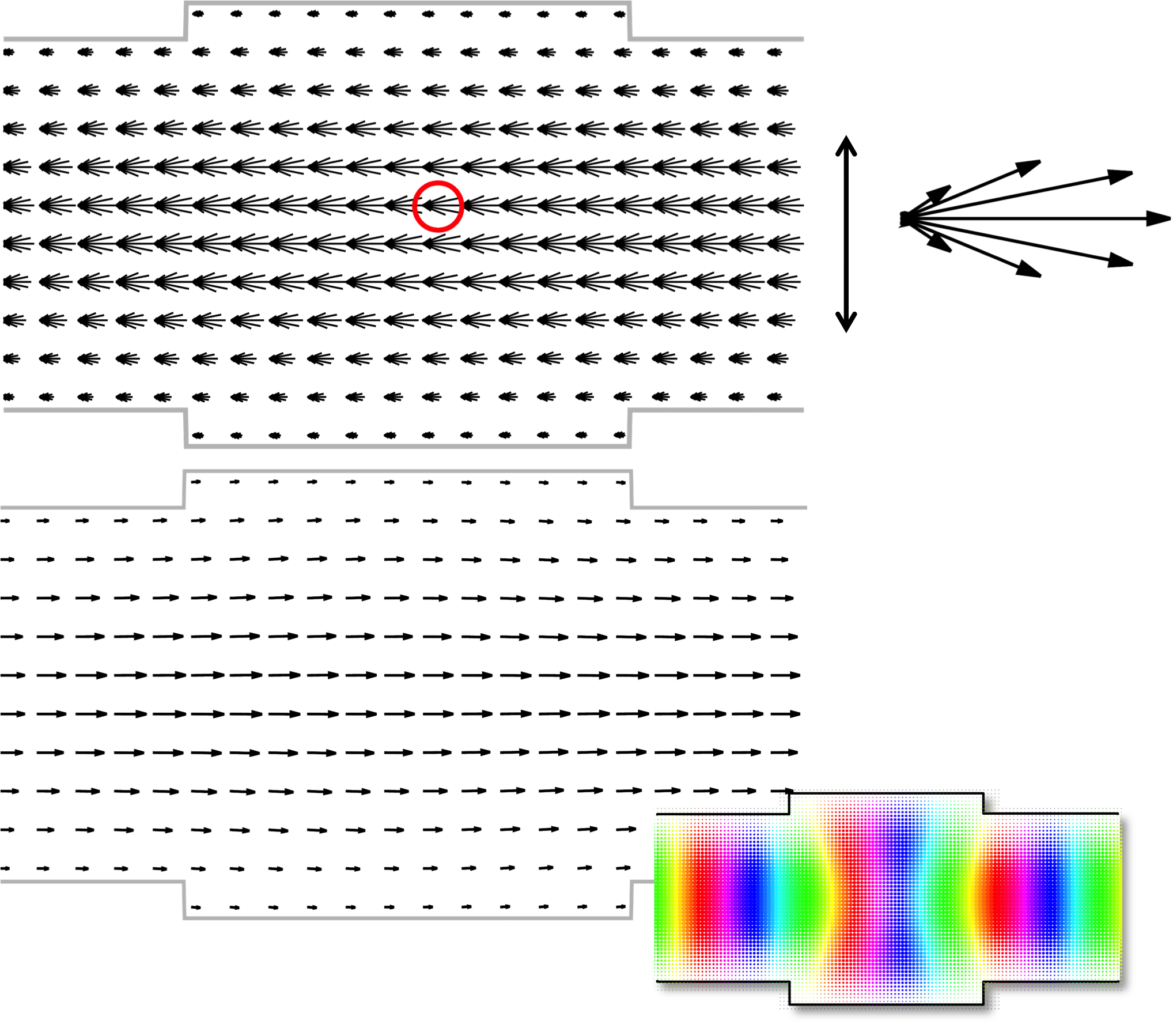}\put(0,87){\figlab{a}}\put(0,48){\figlab{b}}\end{overpic}
\par\end{centering}

\caption{\label{fig:First-Mode}The full Husimi map for lowest propagating
mode wavefunction (see inset) from the waveguide in Fig.~\ref{fig:Sub-Threshold Resonance}
is plotted in (a), at an energy well above resonance ($E=0.02745$
in arbitrary units scaled to Fig.~\ref{fig:Sub-Threshold Resonance}).
The uncertainty for this map is $\Delta k/k=20\%$. At right, a magnified
view of the projection circled in red. In (b), the Husimi flux. This
is the mode which hybridizes with the resonance state to produce Figs.~\ref{fig:Full-Husimis}
and \ref{fig:Above-At-Below}.}
\end{figure}

We can begin to understand these subtle changes by examining the lowest
propagating mode. The full Husimi map far away from resonance, shown
in Fig.~\ref{fig:First-Mode} for a moderate coherent state, corresponds
to the complex plane wave in Fig.~\ref{fig:Plane Waves}. In the\emph{
}Husimi flux, the left-to-right flow appears unchanged within the
central region of the system. The flux operator for this mode, not
shown, is similar. In contrast, the vector-sum and the flux of the
bound state is always zero. So what happens when it interacts with
the lowest propagating mode to produce the resonant state?

\begin{figure}
\begin{centering}
\begin{overpic}[width=1.0\columnwidth]{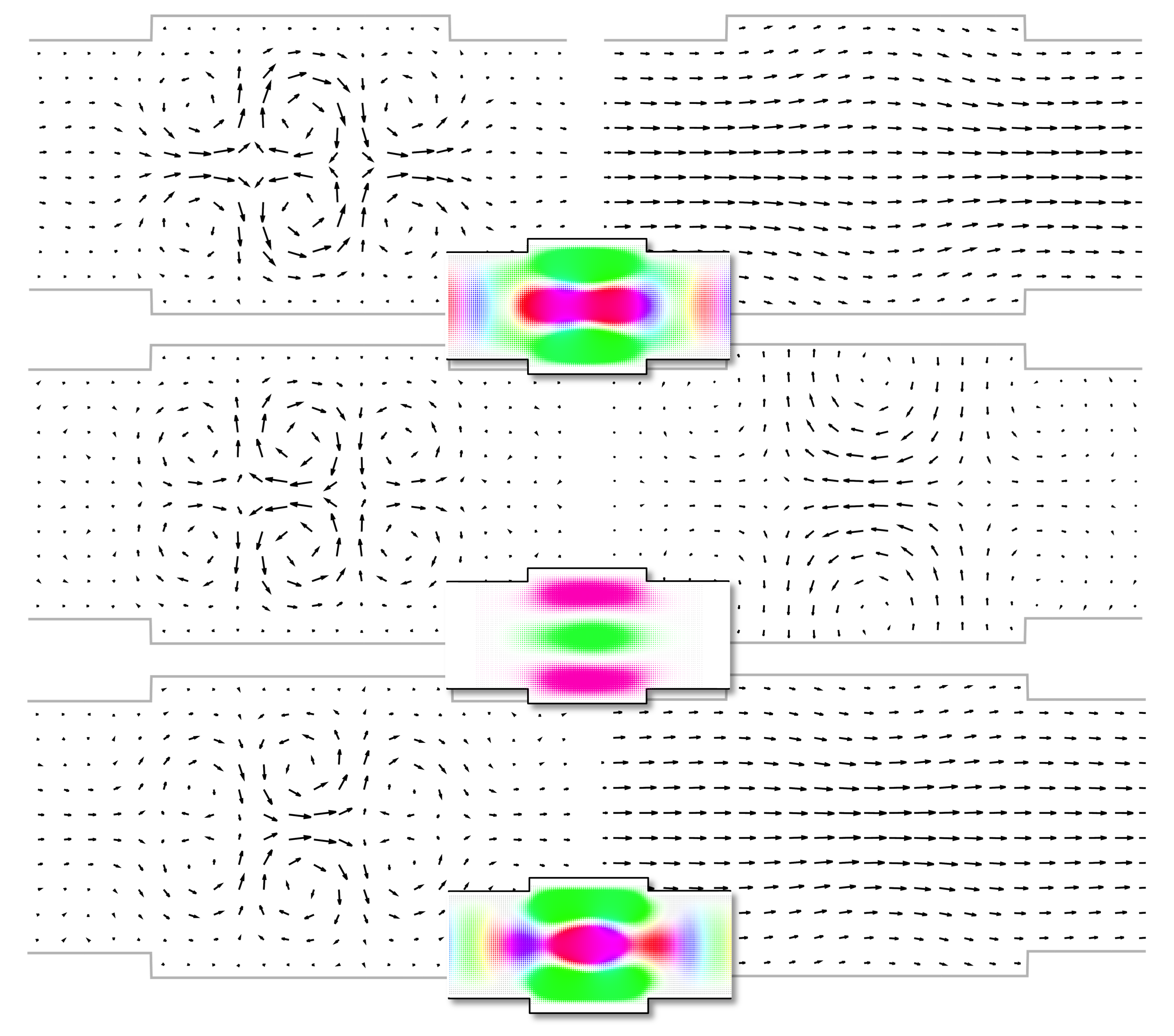}\put(2,87){\figlab{a}}\put(2,59){\figlab{b}}\put(2,31){\figlab{c}}\end{overpic}
\par\end{centering}

\caption{\label{fig:Above-At-Below}The traditional flux (left column) and
the Husimi flux (right column) for the resonance state in Fig.~\ref{fig:Sub-Threshold Resonance}
slightly above resonance (a, $E=E_{\text{res.}}+0.00005$), at resonance
(b, $E=E_{\text{res.}}$) and slightly below resonance (c, $E=E_{\text{res.}}-0.00005$).
The coherent state for the Husimi map corresponds to $\Delta k/k=0.2$.
The transmission function for this mode at each energy corresponds
to $T=0.99$(a), $0.06$(b), and $0.99$(c). Even though the full
Husimi maps at each energy are indistinguishable from Fig.~\ref{fig:Full-Husimis},
their vector additions (Husimi flux) vary substantially. Energies
are in arbitrary units, scaled to Fig.~\ref{fig:Sub-Threshold Resonance}. }
\end{figure}

In Fig.~\ref{fig:Above-At-Below} we address this question by showing
the traditional flux, wavefunction, and the Husimi flux above (a),
at (b), and below (c) resonance. The flux operator is integrated over
a Gaussian kernel corresponding to a coherent state spread of $\Delta k/k=100\%$,
and is visually identical to the Husimi flux with the same coherent
state spread.

In the flux operator, we see the characteristic vortex patterns which
flip above and below resonance, as expected when the bound state shifts
through a phase of $\pi$ over resonance. Moreover, while it is clear
that the presence of the lowest propagating mode is stronger away
from resonance, the wavefunction representation at all three energies
are strongly influenced by the bound state. Similarly, probability
flux is strongly localized in the center of the system, and it is
unclear how the vortices correlate with the fact that transmission
for this mode goes to zero on resonance. 

In the Husimi flux, however, the correlation is obvious. Above and
below resonance, vortices cancel out and leave behind the drift velocity
of the mode. At these energies, the Husimi flux is quite similar to
the lowest propagating mode in Fig.~\ref{fig:First-Mode}, and the
left-to-right flow extends through the semi-infinite leads, although
there are slight changes in the central region.\emph{ }At resonance\emph{,}
however, the vortices no longer interfere to produce flow from left-to-right,
but instead persist as larger vortices across the central region which
counteract the left-to-right flow from the leads, resulting in zero
transmission for this mode. The second-lowest propagating mode (not
shown), which is antisymmetric along the transverse direction, does
not interact with the resonance and maintains full transmission.

At all energies, the full Husimi map shows the simple vertical bouncing
trajectories that are identical to the bound state (Fig.~\ref{fig:Full-Husimis}),
while the left-to-right flow of the lowest propagating mode (Fig.~\ref{fig:First-Mode})
interferes with these paths to produce the residual flux vortices.
The classical dynamics of the resonance therefore indicate a subtle\emph{
}shift in the overall contribution of classical trajectories which
give rise to the resonance. Because the vertical trajectories can
easily cancel each other out, the residual becomes exquisitely sensitive
to the initial conditions of such classical paths, which are determined
solely by the energy of the lowest-propagating mode.

By examining this system using the Husimi flux, which allows us to
adjust the coherent state spread arbitrarily, we can zoom into the
details of the flux operator at small spreads and pull out to larger
drift flows at larger spreads. Important information about the resonance
can be retrieved at all scales, since the flow can be understood by
the slight residuals of the full Husimi projection with different
coherent state spreads. By adding the Husimi projection and the Husimi
flux to the analytical toolset, we can examine the problem from all
angles to construct a more nuanced and complete story.

\subsection{Transport Through Other Geometries and the Nature of Flux Vortices\label{sub:Transport-Through-Other}}

\begin{figure}[t]
\begin{centering}
\begin{overpic}[width=0.95\columnwidth]{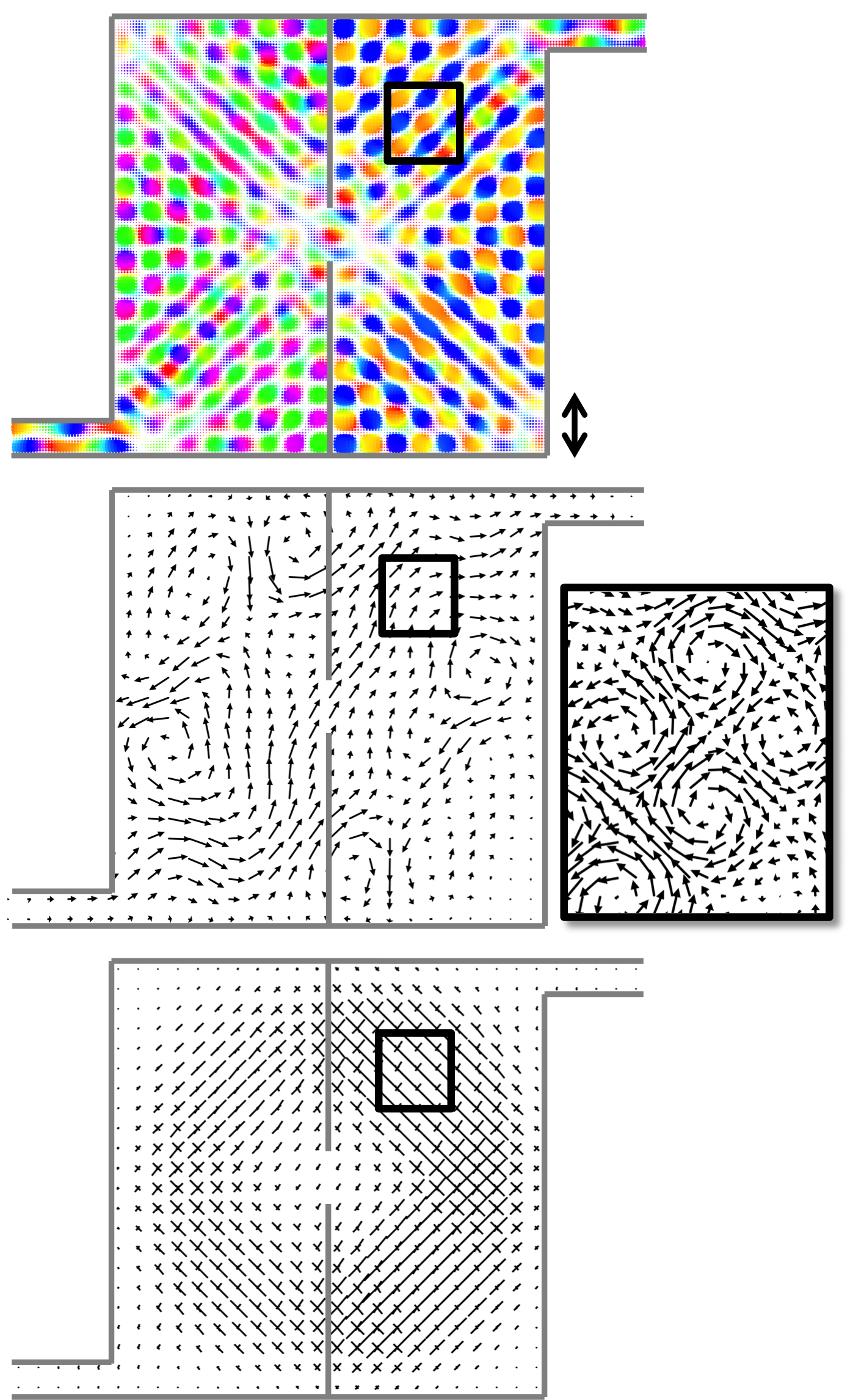}\put(4,97){\figlab{a}}\put(4,63){\figlab{b}}\put(4,29){\figlab{c}}\end{overpic}
\par\end{centering}

\caption{\label{fig:Obstruction-Husimi}A scattering wavefunction associated
with full transmission through the obstructed square with the wavefunction
representation (a), the Husimi flux (b) and multi-modal analysis (c)
for a coherent state spread of $\Delta k/k=10\%$, indicated by the
double arrows. The traditional flux from the part of the system indicated
by the black squares is magnified in the inset.}
\end{figure}

Because the $\sigma$ parameter defines the spatial spread of the
coherent states used to generate a Husimi map, we can use it to reveal
the behavior of the probability flux at arbitrary scales. In Fig.~\ref{fig:Obstruction-Husimi}a,
we show the scattering wavefunction that acts as a mode of unit transmission
in a large square block geometry. This geometry is changed from the
previous subsection so that: 1) Its dimensions are much larger than
the characteristic wavelength at the energies we examine, 2) the leads
are shifted vertically from the center towards the bottom-left and
upper-right, and 3) the center is obstructed to constrain transport
through the central region. As a result, classical paths related to
transport in this system must reflect off the boundaries many times
to propagate from the left to the right lead.
\begin{figure}[t]
\begin{centering}
\includegraphics[width=0.95\columnwidth]{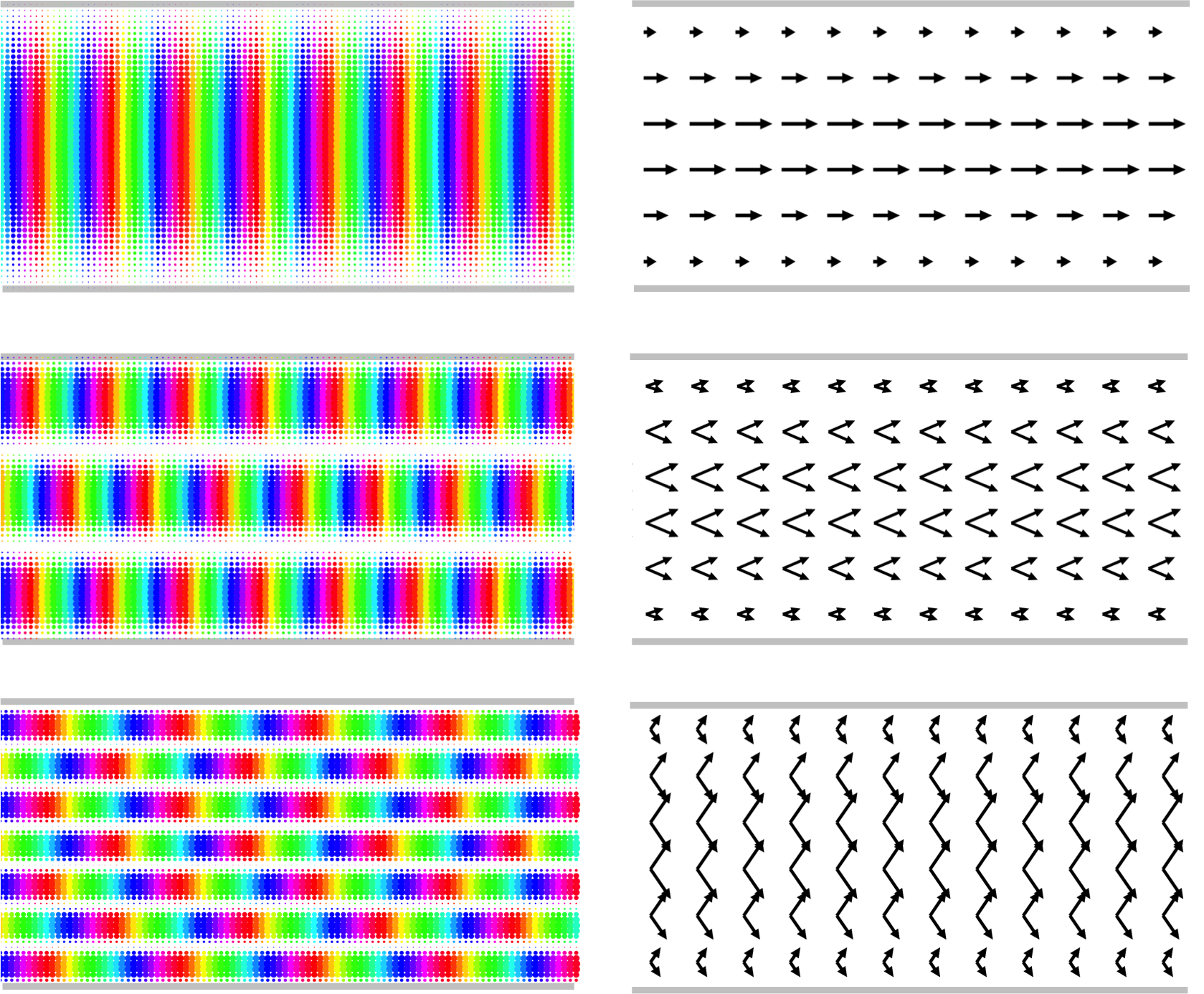}
\par\end{centering}

\caption{\label{fig:Three-Modes}The scattering wavefunction associated with
distinct modes of a wide unperturbed waveguide at left, with the multi-modal
analysis at right. Each mode is associated with a pair of trajectory
angles. As the number of horizontal nodal lines increases, and as
energy increases, these angles become increasingly vertical. }
\end{figure}

In Fig.~\ref{fig:Obstruction-Husimi}, we show the wavefunction,
a magnified view of the traditional flux, the full Husimi flux, and
the multi-modal analysis for this scattering state. In the wavefunction,
nodal lines appear to fall along the $45^{\circ}$ diagonals, which
is corroborated by trajectories favoring those diagonals in the multi-modal
analysis. This arises because all boundary conditions are vertical
or horizontal walls; since each mode of the unperturbed waveguide
leads is associated with a distinct pair of trajectory angles (See
Fig.~\ref{fig:Three-Modes}), the vertical and horizontal walls therefore
reflect all trajectories back onto the same pair rotated at 45 degrees.
At the energy we have selected, the pair of trajectory angles for
the incoming mode are at perfect $45^{\circ}$ diagonals, so that
their rotations from reflecting off the walls also point along the
diagonals, giving rise to strong standing waves.

In Fig.~\ref{fig:Obstruction-Husimi}b, it is clear that transport
occurs primarily through a narrow channel we call the conductance
pathway with the majority of arrows pointing from the lower-left to
the upper-right corners. By comparison, the full traditional flux
map (not shown) is rife with vortices throughout the entire system,
dramatically limiting our ability to identify overall flow. The conductance
pathway does not have to be classical, since it is an aggregate phenomenon
from many other classical trajectories; as a result, it is able to
curve in the bulk without external forces, as seen in the figure.
As the pathway moves against many other perpendicular classical paths
indicated in the multi-modal analysis, pairs of vortices form on either
side, which we show in the inset in Fig.~\ref{fig:Obstruction-Husimi}.
These vortex pairs are a direct analog to those which occur in sub-threshold
resonance as the left-to-right conductance pathway passes through
perpendicular trajectories in the perturbed waveguide (See Fig.~\ref{fig:Above-At-Below}
and the surrounding discussion).

\begin{figure}[t]
\begin{centering}
\begin{overpic}[width=0.95\columnwidth]{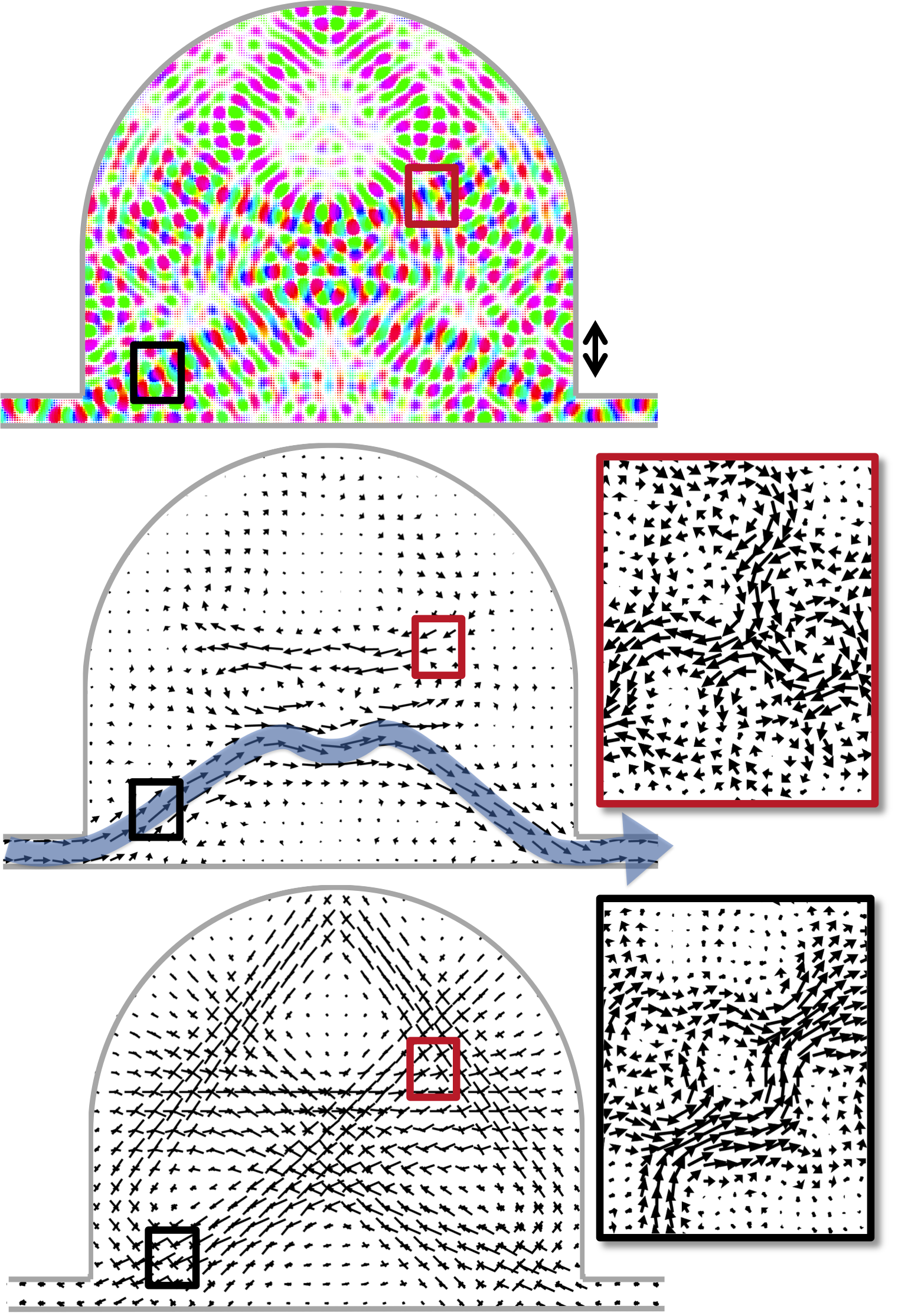}\put(4,97){\figlab{a}}\put(4,63){\figlab{b}}\put(4,29){\figlab{c}}\end{overpic}
\par\end{centering}

\caption{\label{fig:Stadium-Husimi}A scattering wavefunction associated with
full transmission through the half-stadium in the wavefunction representation
(a), the Husimi flux (b) and multi-modal analysis (c) for a coherent
state spread of $\Delta k/k=10\%$, indicated by the double arrows.
The traditional flux from the part of the system indicated by the
black and red squares is magnified in the insets.}
\end{figure}

Fig.~\ref{fig:Stadium-Husimi}, examines a full-transmission scattering
state for a large half-stadium with two leads attached at its sides.
Because scar orbits must self-loop but be otherwise unstable\cite{PhysRevLett.53.1515},
scar states can only participate in transport when the leads attach
at points that are slightly displaced from one of the orbit's reflection
points; otherwise, the classical orbit leaks out the system too quickly.
The wavefunction in Fig.~\ref{fig:Stadium-Husimi} shows strong scarring,
and the multi-modal analysis corroborates the scarring with an identifiable
classical orbit which just misses the leads. 

Like the square device with obstructions in Fig.~\ref{fig:Obstruction-Husimi},
flux also occurs most strongly along a narrow conductance pathway
which, in this case, flows along the bottom of the device while deviating
into the bulk at its middle. In addition, flux vortices occur throughout
the system, making interpretation difficult without applying our methods.
Unlike the square device, however, these vortices no longer form identifiable
pairs. In the stadium state, classical paths do not intersect at $90^{\circ}$
angles, but take on a variety of other oblique angles. As a result,
the vortices take on forms that are consistent with the multi-modal
Husimi map at each intersection. For instance, in the black inset,
there is strong flow from bottom-left to upper-right, with other near-vertical
flows forming vortices, and in the red inset, there are three primary
flows propagating at $60^{\circ}$ to each other, forming the triangular
arrangement of vortices shown.

\section{Conclusions}

We have extended the definition of the flux operator using a new method
for visualizing and analyzing the quantum wavefunction. By using coherent
states as a measurement operator built off of the Husimi projection\cite{Husimi}
and adapting this technique to generate a vector field, we have shown
it is identical to the flux operator for infinitesimal coherent states
(Section \ref{sub:Connection-to-Coherent}). For finite coherent state
spreads, the Husimi projection provides a breakdown of the flux into
contributions from dominant classical paths, which can be retrieved
by processing our results (Section \ref{sub:Multi-Modal-Analysis}).
This technique has proven invaluable for informing a design principle
in quantum systems, since it provides a map of how boundaries affect
individual quantum states (Section \ref{sub:Billiard-Eigenstates}),
as well as the impact of potentials (Section \ref{fig:Ang-Mom-States})
and magnetic fields (Section \ref{sub:Magnetic-Field-States}). Finally,
we have shown its utility for illuminating the many phenomena underlying
resonance when a closed system interacts with an environment (Section
\ref{sub:Sub-Threshold-Resonant-Wavefunct}), while helping to explain
the presence and properties of flux vortices (Section \ref{sub:Transport-Through-Other}).
Because of its ability to contextualize the the flux operator and
identify the primary conductance pathway in large systems, the Husimi
projection is an ideal tool for interpreting quantum conductance simulations.

This paper focuses on two-dimensional systems, since they are ideal
for demonstrating the significant physical intuition that the Husimi
is able to provide. However, its definition is not limited to such
systems. It is equally well suited to three-dimensional systems, and
may be able to provide a significant contribution to interpreting
molecular orbitals, augmenting such technologies as Bader surfaces
analysis\cite{Bader} and local currents\cite{Ratner-Nature}.

\appendix

\section{Deriving the Expectation Value of the Flux\label{sec:Deriving-the-Expectation}}

In this appendix, we show how to derive Eq.~\ref{eq:Flux-Expectation}
from the eigenvalue equation (Eq.~\ref{eq:EigenEq}). We begin by
labeling the excited states of the harmonic oscillator at position
$\vec r_{0}$ oriented along the $i^{\text{th}}$ spatial dimension
\begin{eqnarray}
\braket{\vec r}0 & = & \braket{\vec r}{\vec r_{0},\sigma}\nonumber \\
\braket{\vec r}1 & = & \frac{\vec e_{1}\cdot(\vec r-\vec r_{0})}{\sigma}\braket{\vec r}{\vec r_{0},\sigma}\nonumber \\
\braket{\vec r}2 & = & \sqrt{\frac{1}{2}}\left(\frac{\left(\vec e_{1}\cdot(\vec r-\vec r_{0})\right)^{2}}{\sigma^{2}}-1\right)\braket{\vec r}{\vec r_{0},\sigma}\nonumber \\
 & \vdots
\end{eqnarray}
where $\vec e_{i}$ is the unit vector along the $i^{\text{th}}$
spatial dimension. These states form a complete set in which the flux
operator can be explicitly expressed, using a zero-indexed Hermitian
matrix, as
\begin{equation}
\hat{j}_{\vec r_{0},\sigma,i}=\left(\begin{array}{ccccc}
0 & +i\lambda & 0 & \cdots & 0\\
-i\lambda & 0 & 0 & \cdots & 0\\
0 & 0 & 0 & \cdots & 0\\
\vdots & \vdots & \vdots & \ddots & \vdots\\
0 & 0 & 0 & \cdots & 0
\end{array}\right)
\end{equation}
where $\lambda=\lambda_{\sigma,i,+}=\frac{\hbar}{4m\sigma}$. There
are additional sets of harmonic oscillators orthogonal to the above
states which are centered at points other than $\vec r_{0}$ also
with zero components in the flux matrix

The complete set of eigenstates $\ket{\lambda_{1}},\ket{\lambda_{2}},\ket{\lambda_{3}},\dots$
of the flux operator expressed in terms of excited states of the harmonic
oscillator are 
\begin{equation}
\left(\begin{array}{c}
+1\\
-i\\
0\\
\vdots
\end{array}\right),\left(\begin{array}{c}
+1\\
+i\\
0\\
\vdots
\end{array}\right),\left(\begin{array}{c}
0\\
0\\
1\\
\vdots
\end{array}\right),\cdots
\end{equation}
with eigenvalues $-\lambda$,$\lambda$, and $0$. Measurement by
the flux operator collapses the wavefunction onto one of these eigenstates,
the infinite majority of which are in the degenerate zero-eigenvalue
subspace spanning all excited states of the harmonic oscillator above
$\ket 1$. Only the first two eigenstates, confirmed in Eq.~\ref{eq:Flux-Eigenstate},
yield non-zero flux values, which, as we have already shown, tend
towards positive and negative infinity as $\sigma\rightarrow0$.

When expanding the flux expectation value, we can use the complete
eigenbasis to show that 
\begin{eqnarray}
\braketop{\psi}{\hat{j}_{\vec r_{0},\sigma,i}}{\psi} & = & \braketop{\psi}{\hat{j}_{\vec r_{0},\sigma,i}\sum_{i=1}^{\infty}\ketbra{\vphantom{\sum_{i=1}^{\infty}}\lambda_{i}}{\lambda_{i}}}{\psi}\nonumber \\
 & = & \lambda\left|\braket{\psi}{\lambda_{1}}\right|^{2}-\lambda\left|\braket{\psi}{\lambda_{2}}\right|^{2}.\label{eq:Flux-Eigenstate-Simplified}
\end{eqnarray}
From Eq.~\ref{eq:Flux-Eigenstate}, it can be shown that the contributions
from $\left|\braket{\psi}0\right|^{2}$ and $\left|\braket{\psi}1\right|^{2}$
cancel themselves due to the opposite sign of the eigenvalues, and
only the cross-term $\braket{\psi}0^{\ast}\braket{\psi}1-\braket{\psi}0\braket{\psi}1^{\ast}$
remains. This form is directly related to the commonly-found expression
of the flux at point $\vec r_{0}$ as
\begin{equation}
\vec j_{\vec r_{0}}\left(\Psi(\vec r)\right)=\frac{\hbar}{2mi}\left(\Psi^{*}(\vec r_{0})\boldsymbol{\nabla}\Psi(\vec r_{0})-\Psi(\vec r_{0})\boldsymbol{\nabla}\Psi^{*}(\vec r_{0})\right).
\end{equation}

\section{Uncertainty Propagation for Husimi Vector Addition\label{sub:Uncertainty-Propagation-for}}

When integrating over the available $k$-space in Eq.~\ref{eq:Husimi-Vector},
the resulting Husimi flux vector has lower uncertainty than the individual
terms in the integral, but by how much? Understanding this mathematical
detail is key to appreciating why the Husimi projection is valuable
to extending the flux operator to an operator with defined uncertainty.
Moreover, understanding the behavior of uncertainty propagation in
this integral makes it possible to confidently approximate the result
with a discrete sum, such as the sunbursts in Fig.~\ref{fig:Plane Waves},
offering both visual and computational advantages.

We begin by considering the extreme cases. If the wavevector orientation
remains unchanged for each measurement, summing up identical measurements
has no effect on the final relative uncertainty. On the other hand,
when either the spatial coordinates or the wavevectors are sufficiently
separated, each Husimi vector constitutes an independent measurement;
the uncertainty of the result will reduce by the square root of the
number of measurements. In general, calculations fall in between these
two extremes.

This analysis is concerned with only one dimension, since the variance
along each orthogonal axis can simply be summed. First, the coherent
state is expressed in the momentum basis as 
\begin{eqnarray}
\braket{\vec k}{\vec r_{0},\vec k_{0},\sigma} & = & \left(\frac{2\sigma}{\sqrt{\pi/2}}\right)^{1/2}\nonumber \\
 &  & \times e^{-\sigma^{2}\left(\vec k-\vec k_{0}\right)^{2}+i\left(\vec k-\vec k_{0}\right)\cdot\vec r_{0}}.\label{eq:k-space-coherent}
\end{eqnarray}
Most generally, the Husimi projection in Eq.~\ref{eq:Husimi-Vector}
is the integral of Husimi functions over all of $k$-space. In this
appendix, and in the figures throughout this paper, the integral is
replaced with a finite sum of test wavevectors $\left\{ \vec k_{i}\right\} $
which satisfy the dispersion relation at a particular energy. 

The variance of the integral in Eq.~\ref{eq:Husimi-Vector} can be
obtained by building on intuition about coherent states. It is well-known
that the $k$-space variance of the coherent state can be simply derived
by integrating the coherent state probability amplitude over $k$-space,
weighting the integrand by $\left(\vec k-\vec k_{0}\right)^{2}$.
Using the notation in Eq.~\ref{eq:k-space-coherent}, this gives
$\sigma_{k}^{2}=\frac{1}{4\sigma_{x}^{2}}$ yielding the familiar
relation $\sigma_{x}\sigma_{k}=\frac{1}{2}$. This can be thought
of in the Husimi formulation as a statistical result where the quantity
$\sigma_{k}$ is the variance of each individual term in the Husimi
vector summation. In this formulation, the variable is the wavevector
and the probability function is the probability amplitude of the coherent
state. Because the probability function is complex, we have to take
the absolute sum squared. 

Factoring in more than one Husimi function into the Husimi projection
results in the expression 
\begin{equation}
\frac{2\sigma}{\sqrt{\pi/2}}\int_{-\infty}^{\infty}\left|\sum_{i}\left(k-k_{i}\right)e^{-\sigma^{2}\left(k-k_{i}\right)^{2}+i\left(k-k_{i}\right)x_{0}}\right|^{2}dk,
\end{equation}
where the set $\left\{ \vec k_{i}\right\} $ are the set of test wavevectors,
projected onto the given axis, $x_{0}$ is the spatial point being
tested, and $\sigma$ is the chosen spatial Gaussian spread. Setting
the coherent states to the same phase at their centers, $x_{0}=0$,
and the above integral can be evaluated to return
\begin{equation}
\sigma_{k}^{2}=\frac{1}{4\sigma^{2}}\left(N+2\sum_{i,j>i}e^{-\frac{\sigma^{2}}{2}\left(k_{i}-k_{j}\right)^{2}}\left(1-\sigma^{2}\left(k_{i}-k_{j}\right)^{2}\right)\right).\label{eq:Uncertainty}
\end{equation}

Already it is possible test this result against intuition. If each
wavevector is identical, then $k_{i}-k_{j}=0$ and the sum of $N$
measurements results in the uncertainty $\sigma_{k}^{2}=\frac{N^{2}}{4\sigma^{2}}$
which would provide no reduction of relative uncertainty. For large
values of $\left|k_{i}-k_{j}\right|\gg\sigma$, the exponential term
will overwhelm the quadratic term and the uncertainty becomes $\sigma_{k}^{2}=\frac{N}{4\sigma^{2}}$,
a reduction in the relative uncertainty of $\sqrt{N}$. 

\begin{figure}
\begin{centering}
\includegraphics[angle=-90,width=0.85\columnwidth]{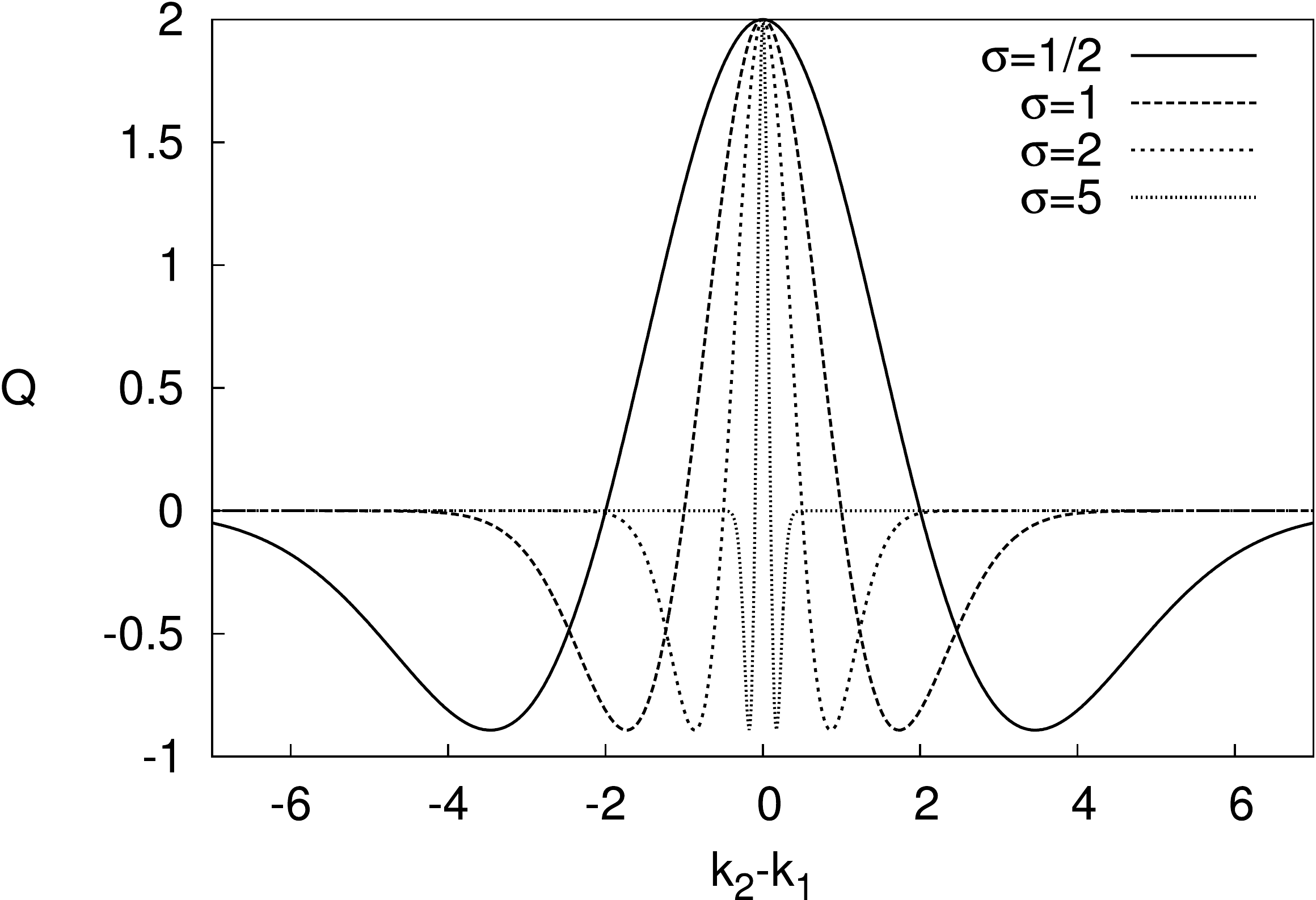}
\par\end{centering}

\caption{\label{fig:Two-Vectors}The second term in Eq.~\ref{eq:Uncertainty}
is plotted for the addition of two vectors in the Husimi projection.
This term represents the covariance between the two vectors, and is
bounded above by $2$ and below by $-\frac{4}{e^{3/2}}$ for all choices
of $\sigma$.}
\end{figure}

Perhaps most surprising about Eq.~\ref{eq:Uncertainty} is that the
second term, which quantifies the covariance between the two measurements,
can actually be negative. What are its bounds? Fig.~\ref{fig:Two-Vectors}
plots the quantity $Q(k_{1},k_{2},\sigma)=2e^{-\frac{\sigma^{2}}{2}\left(k_{2}-k_{1}\right)^{2}}\left(1-\sigma^{2}\left(k_{2}-k_{1}\right)\right)$,
showing that a minimum value of $-\frac{4}{e^{3/2}}\approx-0.893$
is achieved at $\left|k_{2}-k_{1}\right|=\sqrt{3}/\sigma$. Every
value of $\left|k_{2}-k_{1}\right|$ beyond which $Q$ goes through
zero has achieved nearly independent measurements, which is found
at $\left|k_{2}-k_{1}\right|=\sigma^{-1}$.

\begin{figure}
\begin{centering}
\includegraphics[angle=-90,width=0.85\columnwidth]{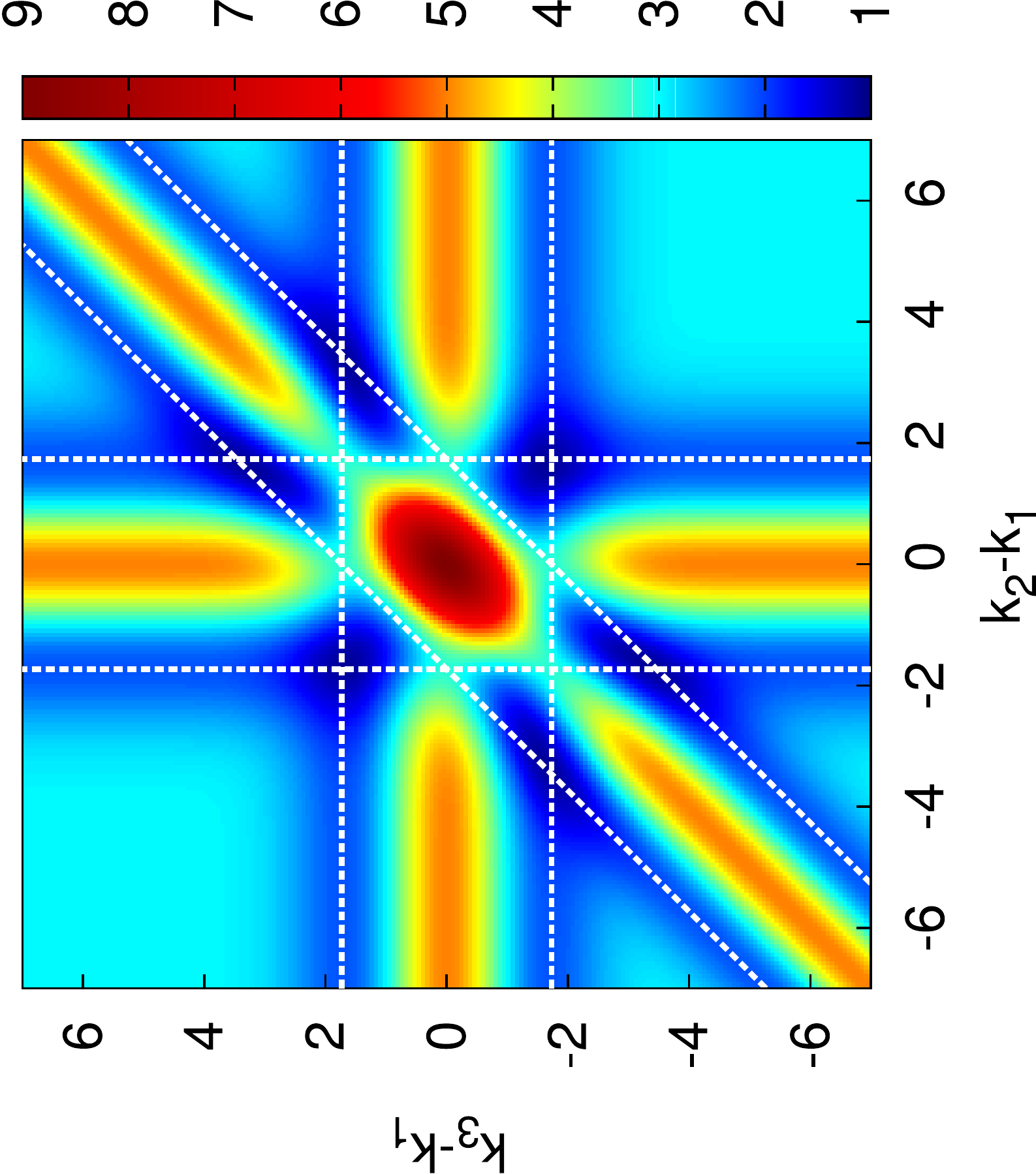}
\par\end{centering}

\caption{\label{fig:Three-Vectors}The uncertainty that results from summing
three vectors of a Husimi projection, as written in Eq.~\ref{eq:Uncertainty}
is plotted. The uncertainty is bounded above by $9/4\sigma^{2}$ and
below by $\sim1.017/4\sigma^{2}$. The dashed white lines indicate
local minima that result from spacing each pair of vectors by $\sqrt{3}/\sigma$,
which would give a minimum uncertainty for two-vector addition (see
Fig.~\ref{fig:Two-Vectors}). }
\end{figure}

The terms in Eq.~\ref{eq:Uncertainty} suggest that when more and
more vectors are added the uncertainty can be reduced arbitrarily
by setting the correct separations between the test wavevectors. It
even suggests that for three or more vectors we could possibly produce
results with negative uncertainty, but intuitively that cannot be
possible. To appreciate why from an analytical perspective, Fig.~\ref{fig:Three-Vectors}
plots the results of $\sigma_{k}^{2}$ for the addition of three wavevectors.
The minima that occur from maximizing the separation between each
pair of wavevectors is indicated by the white dashed lines. At the
center of the graph, a peak exists at $\sigma_{k}^{2}=9/4\sigma^{2}$,
which falls to $3/4\sigma^{2}$ for areas beyond the area bounded
by the white dashed lines, consistent with earlier observations. There
is also a minimum (positive) uncertainty which arises from the fact
that the separation between all pairs of points on a line cannot be
equal. In Fig.~\ref{fig:Three-Vectors} this is evidenced by the
fact that there are no points where three dashed lines intersect.
For two vectors the minimum occurs at $\sigma_{k}^{2}\approx0.981/4\sigma^{2}$,
for three $\sigma_{k}^{2}\approx1.017/4\sigma^{2}$ and for four $\sigma_{k}^{2}\approx1.036/4\sigma^{2}$.
We can generalize and state that for $N_{\text{min}}$ vectors that
fall on separate minima, the uncertainty of their sum will be $\sigma_{k}\approx\frac{1}{2N_{\text{min}}\sigma}$.

Moreover, even if vectors are added that do not fall on the uncertainty
minima in Figs.~\ref{fig:Two-Vectors} and \ref{fig:Three-Vectors},
they will have a negligible impact on the total relative uncertainty.
So no matter how many vectors contribute to the sum, only the vectors
on the minima will reduce the relative uncertainty, making the key
quantity not the total number of vectors that are added, but the number
that have sufficient separation to fall on the uncertainty minima. 

How many vectors is this? We know, for instance, that this minimum
occurs when the maximum number of vector pairs have a separation near
$\sqrt{3}/\sigma$, and that this is likely to occur when they are
evenly spaced on a line at that separation. Thus we propose that the
number of vectors that can fall on the minima is given by $N_{\text{min}}=\text{floor}\left(2k(E)\sigma/\sqrt{3}\right)$,
and using $\hbar k=\sqrt{2mE}$, we can rewrite this as $N_{\text{min}}=\text{floor}\left(\sigma\sqrt{\frac{8mE}{3\hbar^{2}}}\right)$.
Substituting this value results in the proportionalities 
\begin{equation}
\Delta k/k\propto\frac{1}{N_{\text{min}}\sigma}\propto\left(\frac{\sigma}{\hbar}\sqrt{mE}\right)^{-1}.
\end{equation}
 This makes sense intuitively: the relative uncertainty of a finely
sampled Husimi vector addition goes down with larger $\sigma$ and
energy. 

This result deepens the connection between the flux operator and the
Husimi function for small $\sigma$, since for very small coherent
states, the uncertainty minima, which are separated by $\sigma^{-1}$,
grow increasingly far apart. There is only a finite range of wavevectors
which satisfy the dispersion relation at a given energy, meaning that
as the coherent states get smaller, fewer and fewer samples in $k$-space
minimize the uncertainty. In fact, at the extreme limit of $\sigma\rightarrow0$,
the uncertainty cannot be minimized beyond a single measurement in
each orthogonal direction, indicating that results for these small
coherent states have undefined uncertainty, just like the flux operator.
We corroborate this result using a different proof in Eq.~\ref{eq:Flux-Hus-Corr}.

\section{The Hamiltonian\label{sec:The-Hamiltonian}\label{sec:Magnetic-Fields}}

Numerical simulations in this paper use a free-particle Hamiltonian
$H=-\frac{p^{2}}{2m}+U(\vec r)$ sampled on a square grid with spacing
$a$ and where $U(\vec r)=0$ at all points unless otherwise stated.
This Hamiltonian can be expressed in more familiar language by using
the tight-binding approximation. In this approximation, the effective
mass envelope function Hamiltonian becomes $H=\sum_{i}\epsilon_{i}\mathbf{a}_{i}^{\dagger}\mathbf{a}_{i}-t\sum_{\left\langle ij\right\rangle }\mathbf{a}_{i}^{\dagger}\mathbf{a}_{j}$
where $\mathbf{a}_{i}$ is the annihilation operator for the $i^{\text{th}}$
grid point, $\epsilon_{i}$ is the energy of the system plus the disorder
potential, and the set $\left\langle ij\right\rangle $ cycles through
all nearest-neighbor pairs. This gives the hopping term $t=\frac{\hbar^{2}}{2ma^{2}}$
and $\epsilon_{i}=4t+U_{i}$, where $U_{i}$ is everywhere zero unless
otherwise stated.

Sec.~\ref{sub:Magnetic-Field-States} uses the Peierls substitution\cite{peierls}
to incorporate magnetic fields, using the language of the tight-binding
model. In this model, the magnetic field contributes a phase to the
hopping potential $t$: 
\begin{equation}
t_{ij}=t\exp\left[i\phi\right],\phi=q\vec A\cdot(\vec r_{i}-\vec r_{j})/\hbar,\label{eq:mag_phase}
\end{equation}
where $\vec r_{i}$ is the position vectors of the site corresponding
to the $i^{\text{th}}$ column of the Hamiltonian, $\hbar$ is Planck's
constant, and $q$ is the electron charge. Calculations in this paper
assume that the magnetic field is perpendicular to the plane on which
the system sits and is bounded by a cylinder centered on the system's
center. The radius of this column is chosen to be greater than the
size of the system. Accordingly, the gauge of the magnetic potential
for an out-of-plane magnetic field is defined such that

\begin{equation}
\vec A(\vec r)=\frac{\mathbf{e}_{\theta}}{2\pi r}\int B_{z}dxdy,
\end{equation}
where the integral is over a disc centered on the origin and limited
by radius $r$.

The cyclotron radius can be determined by the relation
\begin{equation}
r=\frac{\hbar k}{B_{0}q}.
\end{equation}
For a free particle, $\hbar k=\sqrt{2mE}$, giving 
\begin{equation}
\frac{r}{a}=\frac{\sqrt{2mE}}{B_{0}qa^{2}}.
\end{equation}
This means that at $E=0.2\frac{\hbar^{2}}{ma^{2}}$, the energy used
in Sec.~\ref{sub:Magnetic-Field-States}, a magnetic field strength
of $B_{0}=2\times10^{-3}\frac{\hbar}{qa^{2}}$ is sufficient to produce
a cyclotron radius that is $2/3$ of the system radius. This relation
is used to predict the cyclotron radius for all calculations in this
paper.

\section{Scattering Wavefunctions\label{sec:Calculating-the-Wavefunctions}}

Diagonalizing the Hamiltonian to examine eigenstates of a closed system
is straightforward. Sec.~\ref{sub:Sub-Threshold-Resonant-Wavefunct},
however, examines an open system in a standard ballistic conductance
calculation. The numerical Green's function formalism is used to obtain
the scattering wavefunction for these calculations, for which modern
implementations are outlined in several texts\cite{datta,datta2,ferrybook}
. In this formalism, the Hamiltonian is divided into a left-lead,
central region, and right-lead projections

\begin{equation}
H=\left(\begin{array}{ccc}
H_{L} & V_{LC} & 0\\
V_{LC}^{\dagger} & H_{C} & V_{RC}\\
0 & V_{RC}^{\dagger} & H_{R}
\end{array}\right).
\end{equation}
The semi-infinite Green's function at the surface of each lead is
calculated using the Lopez-Sanchos method, written it as $g_{L,R}(E)$
for the left ($L$) and right ($R$) leads, which are both equal.
To compute the Green's function for the infinite system within the
device $G(E)$ the semi-infinite surface Green's functions $g(E)$
for each lead\cite{sancho1,sancho2} are first computed and matched
to the surface Green's function of the device region, using the numerical
technique outlined in Mason et al.\cite{mason}.

The coupling matrix for the left lead to the central region is then
defined by $\Gamma_{L}(E)=2\text{Im}\left[V_{LC}^{\dagger}g_{L}(E)V_{LC}\right]$.
This results in a density matrix of coherent scattering wavefunctions
$\rho=G\Gamma_{L}G$ where we have dropped the implicit energy dependence.
Each coherent scattering wavefunction in the system can be obtained
by diagonalizing $\rho$. Associated with each eigenvector of $\rho$
will be an eigenvalue equal to the likelihood of measuring the wavefunction
within the system. Since there are generally more basis sets within
the central region than modes available to the system through the
semi-infinite leads, the vast majority of the eigenvalues will be
zero, and the number of non-zero eigenvalues will be equal to the
number of modes available to the system at the given energy. This
number determines the maximum transmission across the central region. 

Since a resonant state ``traps'' the wavefunction at a specific
energy, it creates a striking peak in the density of states (see Fig.~\ref{sub:Sub-Threshold-Resonant-Wavefunct}).
As a result, the resonant state can be easily identified among the
eigenvectors of the density matrix since it will be associated with
the largest eigenvalue near the resonance energy. When discussing
resonant wavefunctions, it will be assumed that we are using a density
matrix near the resonance energy and examining the eigenvector associated
with the largest eigenvalue (and measurement probability) at that
energy. This makes it possible to distinguish the resonant wavefunction
from other modes which are propagating through the system but are
unaffected by the resonance.
\begin{acknowledgments}
This research was conducted with funding from the Department of Energy
Computer Science Graduate Fellowship program under Contract No. DE-FG02-97ER25308.
MFB and EJH were supported by the Department of Energy, office of
basic science (grant DE-FG02-08ER46513).
\end{acknowledgments}
\bibliographystyle{unsrt}

\begin{thebibliography}{10}

\bibitem{PhysRevLett.53.1515}
Eric~J. Heller.
\newblock Bound-state eigenfunctions of classically chaotic hamiltonian
  systems: Scars of periodic orbits.
\newblock {\em Phys. Rev. Lett.}, 53(16):1515--1518, Oct 1984.

\bibitem{Husimi}
K.~Husimi.
\newblock Some formal properties of the density matrix.
\newblock {\em Proc. Phys. Math. Soc. Jpn.}, 22:264--314, 1940.

\bibitem{Mason-PRL}
Douglas~J. Mason, Mario~F. Borunda, and Eric~J. Heller.
\newblock A semiclassical interpretation of probability flux.
\newblock {\em arXiv}, 2012.

\bibitem{measurement3}
W.~Gale, E.~Guth, and G.~T. Trammell.
\newblock Determination of the quantum state by measurements.
\newblock {\em Phys. Rev.}, 165:1434--1436, Jan 1968.

\bibitem{Measurement-schrod}
Yakir Aharonov and Lev Vaidman.
\newblock Measurement of the schr{\"o}dinger wave of a single particle.
\newblock {\em Physics Letters A}, 178:38 -- 42, 1993.

\bibitem{measurement2}
M.~Daumer, D.~D{\"u}rr, S.~Goldstein, and N.~Zanghi.
\newblock On the quantum probability flux through surfaces.
\newblock {\em Journal of Statistical Physics}, 88:967--977, 1997.

\bibitem{hellerleshouches}
E.~J. Heller.
\newblock Wavepacket dynamics and quantum chaology.
\newblock In M.~J. Giannoni, A.~Voros, and J.~Zinn-Justin, editors, {\em
  Proceedings of the 1989 Les Houches Summer School on ``Chaos and Quantum
  Physics''}, pages 546--663, North-Holland, 1989. Elsevier Science Publishers
  B.V.

\bibitem{Husimi-map-old}
P.W. O'Connor and S.~Tomsovic.
\newblock The unusual nature of the quantum baker's transformation.
\newblock {\em Annals of Physics}, 201(1):218--264, 1991.

\bibitem{Husimi-map-old2}
M.S. Child, G.~Bruun, and R.~Paul.
\newblock Short time quantum phase space dynamics at a 1:2 fermi resonance.
\newblock {\em Chemical Physics}, 190:373 -- 380, 1995.
\newblock Overtone Spectroscopy and Dynamics.

\bibitem{ARPES-Review}
Andrea Damascelli.
\newblock Probing the electronic structure of complex systems by arpes.
\newblock {\em Physica Scripta}, 2004(T109):61, 2004.

\bibitem{gutzwiller-chaos}
M.C. Gutzwiller.
\newblock {\em Chaos in classical and quantum mechanics}.
\newblock Interdisciplinary applied mathematics. Springer-Verlag, 1990.

\bibitem{heller-more-husimi}
L.~Kaplan and E.~J. Heller.
\newblock Measuring scars of periodic orbits.
\newblock {\em Phys. Rev. E}, 59(6):6609--6628, Jun 1999.

\bibitem{Heller-billiard-with-Husimi}
W.~E. Bies, L.~Kaplan, M.~R. Haggerty, and E.~J. Heller.
\newblock Localization of eigenfunctions in the stadium billiard.
\newblock {\em Phys. Rev. E}, 63:066214, May 2001.

\bibitem{circ-well-quan-class-corr}
R.~W. Robinett.
\newblock Visualizing the solutions for the circular infinite well in quantum
  and classical mechanics.
\newblock {\em American Journal of Physics}, 64(4):440--446, 1996.

\bibitem{Bunimovich}
L.~A. Bunimovich.
\newblock On the ergodic properties of some billiards.
\newblock {\em Funct. Anal. App.}, 8:73--74, 1974.

\bibitem{chaology0}
M.~V. Berry.
\newblock The bakerian lecture, 1987: Quantum chaology.
\newblock {\em Proceedings of the Royal Society of London. A. Mathematical and
  Physical Sciences}, 413(1844):183--198, 1987.

\bibitem{Heller-billiard4}
Patrick~W. O'Connor and Eric~J. Heller.
\newblock Quantum localization for a strongly classically chaotic system.
\newblock {\em Phys. Rev. Lett.}, 61:2288--2291, Nov 1988.

\bibitem{chaology2}
Michael~V Berry.
\newblock Quantum chaology, not quantum chaos.
\newblock {\em Physica Scripta}, 40(3):335--336, 1989.

\bibitem{Heller-billiard}
S.~Sridhar and E.~J. Heller.
\newblock Physical and numerical experiments on the wave mechanics of
  classically chaotic systems.
\newblock {\em Phys. Rev. A}, 46:R1728--R1731, Aug 1992.

\bibitem{Heller-billiard2}
Steven Tomsovic and Eric~J. Heller.
\newblock Long-time semiclassical dynamics of chaos: The stadium billiard.
\newblock {\em Phys. Rev. E}, 47:282--299, Jan 1993.

\bibitem{billard-scars}
Fernando~P. Simonotti, Eduardo Vergini, and Marcos Saraceno.
\newblock Quantitative study of scars in the boundary section of the stadium
  billiard.
\newblock {\em Phys. Rev. E}, 56:3859--3867, Oct 1997.

\bibitem{Heller-billiard5}
Alex Barnett, Doron Cohen, and Eric~J. Heller.
\newblock Deformations and dilations of chaotic billiards: Dissipation rate,
  and quasiorthogonality of the boundary wave functions.
\newblock {\em Phys. Rev. Lett.}, 85:1412--1415, Aug 2000.

\bibitem{Deformation-Chaotic-Billiard}
Alex Barnett, Doron Cohen, and Eric~J. Heller.
\newblock Deformations and dilations of chaotic billiards: Dissipation rate,
  and quasiorthogonality of the boundary wave functions.
\newblock {\em Phys. Rev. Lett.}, 85:1412--1415, Aug 2000.

\bibitem{Deformation-Cavity}
Doron Cohen, Alex Barnett, and Eric~J. Heller.
\newblock Parametric evolution for a deformed cavity.
\newblock {\em Phys. Rev. E}, 63:046207, Mar 2001.

\bibitem{Feshbach1958357}
Herman Feshbach.
\newblock Unified theory of nuclear reactions.
\newblock {\em Annals of Physics}, 5(4):357 -- 390, 1958.

\bibitem{Bader}
R~F~W Bader.
\newblock {\em Atoms in Molecules: a Quantum Theory}.
\newblock New York: Oxford University Press, 1990.

\bibitem{Ratner-Nature}
Gemma~C. Solomon, Carmen Herrmann, Thorsten Hansen, Vladimiro Mujica, and
  Mark~A. Ratner.
\newblock Exploring local currents in molecular junctions.
\newblock {\em Nat Chem}, 2(3):223--228, 03 2010.

\bibitem{peierls}
R.~Peierls.
\newblock Zur theorie des diamagnetismus von leitungselektronen.
\newblock {\em Zeitschrift f{\"u}r Physik A Hadrons and Nuclei}, 80:763--791,
  1933.
\newblock 10.1007/BF01342591.

\bibitem{datta}
S.~Datta.
\newblock {\em Electronic Transport in Mesoscopic Systems}.
\newblock Cambridge University Press, Cambridge, 1997.

\bibitem{datta2}
S.~Datta.
\newblock {\em Quantum transport: atom to transistor}.
\newblock Cambridge University Press, 2005.

\bibitem{ferrybook}
D.K. Ferry and S.M. Goodnick.
\newblock {\em Transport in nanostructures}.
\newblock Cambridge Studies in Semiconductor Physics and Microelectronic
  Engineering. Cambridge University Press, 1999.

\bibitem{sancho1}
M.~P. L{\'o}pez-Sancho and J.~Rubio.
\newblock Quick iterative scheme for the calculation of transfer matrices:
  application to mo (100).
\newblock {\em J. Phys. F.: Met. Phys.}, 14:1205--1215, 1984.

\bibitem{sancho2}
M.~P. L{\'o}pez-Sancho and J.~Rubio.
\newblock Highly convergent schemes for the calculation of bulk and surface
  green functions.
\newblock {\em J. Phys. F.: Met. Phys.}, 15:851--858, 1985.

\bibitem{mason}
Douglas~J. Mason, David Prendergast, Jeffrey~B. Neaton, and Eric~J. Heller.
\newblock Algorithm for efficient elastic transport calculations for arbitrary
  device geometries.
\newblock {\em Phys. Rev. B}, 84:155401, Oct 2011.

\end{thebibliography}

\end{document}